\documentclass[a4paper, amsfonts, amssymb, amsmath, reprint, showkeys, nofootinbib, twoside]{revtex4-1}
\usepackage[english]{babel}
\usepackage[utf8]{inputenc}
\usepackage[colorinlistoftodos, color=green!40, prependcaption]{todonotes}

\usepackage{hyperref}
\usepackage{amsmath}

\usepackage{mathtools}
\usepackage{physics}
\usepackage{xcolor}
\usepackage{graphicx}
\usepackage[left=23mm,right=13mm,top=35mm,columnsep=15pt]{geometry} 
\usepackage{adjustbox}
\usepackage{placeins}
\usepackage[T1]{fontenc}
\usepackage{lipsum}
\usepackage{csquotes}

\usepackage{amsthm}
\usepackage{mathtools}
\usepackage{physics}
\usepackage{xcolor}
\usepackage{graphicx}
\usepackage[left=23mm,right=13mm,top=35mm,columnsep=15pt]{geometry} 
\usepackage{adjustbox}
\usepackage{placeins}
\usepackage[T1]{fontenc}
\usepackage{lipsum}
\usepackage{csquotes}

\usepackage{mathrsfs}

\usepackage{subcaption}
\begin{document}

%\title{Hybrid Gottesman–Kitaev–Preskill entanglement from single-photon resources}
\title{Linear-optical generation of hybrid GKP entanglement from small-amplitude cat states}

\author{Shohei Kiryu$^{1}$}
\email{kiryu.opt@keio.jp}
\author{Yohji Chin$^{2}$}
\author{Masahiro Takeoka$^{1,3}$}
\author{Kosuke Fukui$^{2}$}

\affiliation{$^{1}$ Department of Electronics and Electrical Engineering, Keio University, 3-14-1 Hiyoshi, Kohoku-ku, Yokohama 223-8522, Japan}
\affiliation{$^{2}$ Department of Applied Physics, School of Engineering, The University of Tokyo, 7-3-1 Hongo, Bunkyo-ku, Tokyo 113-8656, Japan}
\affiliation{$^{3}$ National Institute of Information and Communications Technology (NICT), Koganei, Tokyo 184-8795, Japan}

\date{\today} % Leave empty to omit a date

\begin{abstract}
    Hybrid bosonic codes combining bosonic codes with photon states offer a promising pathway for fault-tolerant quantum computation. However, the efficient generation of such states in optical setups remains technically challenging due to the requirement for complex non-Gaussian resources. In this paper, we propose a novel scheme to efficiently generate hybrid entangled states between a GKP qubit and a photon-number state using small-amplitude cat states as the primary resource. 
    We apply a breeding process using small-amplitude cat states to increase the non-Gaussianity of the input states. This method requires only linear optical elements and homodyne measurements.
    Furthermore, we demonstrate that this protocol can be extended to generate hybrid qudit states. This scheme has the potential to provide a resource-efficient and experimentally attractive route toward implementing hybrid quantum error correction.

\end{abstract}

\maketitle

\section{INTRODUCTION}
    Quantum error correction encodes logical information into an enlarged physical Hilbert space~\cite{shor1995scheme,kitaev2003fault}. Discrete-variable quantum error correction codes typically require many physical qubits to realize a single logical qubit, which leads to substantial overhead in both resource scaling and logical operations~\cite{horsman2012surface}.
    
    Bosonic codes provide a hardware-efficient alternative by encoding logical information in the infinite-dimensional Hilbert space of a single harmonic oscillator \cite{ralph2003quantum,Gottesman2001GKP,michael2016new}. 
    The Gottesman–Kitaev–Preskill (GKP) code~\cite{Gottesman2001GKP} has emerged as a particularly promising realization. It offers pronounced robustness against optical loss \cite{Fukui2017GkpAnalog,Albert2018GkpLoss}. Approximate GKP states can be generated by interfering cat states on a beam splitter followed by homodyne measurement~\cite{Weigand2018Breed}.

    However, the implementation of single-mode bosonic codes still entails stringent hardware requirements. For instance, achieving sufficient error-correction capabilities in cat codes~\cite{ralph2003quantum} necessitates the generation of cat states with large amplitudes~\cite{lund2008fault}. 
    Similarly, realizing fault-tolerant operation with GKP codes requires exceedingly high squeezing levels~\cite{fukui2018high,bourassa2021blueprint,larsen2021fault,noh2022low,fukui2023high}. 
    Because highly macroscopic quantum features are inherently susceptible to environmental noise and optical loss during generation, realizing the highly demanding parameter regimes necessary for fault tolerance in single-mode codes constitutes a substantial experimental obstacle~\cite{su2019conversion,bourassa2021blueprint,su2022universal,takase2023gottesman,endo2025high,larsen2025integrated,haussler2025long}.
    
    To address this challenge, hybrid bosonic codes are emerging as a promising approach~\cite{ andersen2015hybrid,lee2013near,jeong2014generation,lee2015nearly,omkar2020resource,omkar2021highly,lee2024fault,bose2024long,fukui2024resource,kiryu2025linear,haussler2025long,bera2025long}. 
    These codes combine a bosonic mode with a discrete two-level system, such as polarization or photon number states. They enable continuous and discrete variables to complement each other: discrete systems offer accessible operations, while bosonic modes provide hardware-efficient encoding. Crucially, hybrid architectures substantially improve error-correction performance and relax demanding hardware constraints, notably the squeezing levels required for fault tolerance~\cite{lee2013near,fukui2024resource}. These robust properties also make hybrid states highly suitable for quantum communication networks. In particular, owing to their ability to combine robust continuous-variable transmission with experimentally accessible discrete-variable measurements, they have been proposed as crucial resources for realizing long-distance quantum communication, such as in quantum repeaters~\cite{haussler2025long}.
    
    While hybrid states require significantly less demanding hardware parameters for fault tolerance, their initial preparation in optical platforms remains a major experimental hurdle~\cite{kwon2014generation,jeong2014generation,fukui2024resource,kiryu2025linear,haussler2025long}. 
    Standard generation protocols employ interference of Gaussian states followed by conditional projective measurements. Their reliance on multiple probabilistic non-Gaussian resources~\cite{dakna1997generating,lund2004conditional,takase2023gottesman,aghaee2025scaling,solodovnikova2025loss,hanamura2025beyond} results in exponentially small success probabilities and correspondingly low generation rates~\cite{kwon2014generation,jeong2014generation}.
    To overcome this limitation, recent proposals generate target single-mode bosonic states by breeding single photons. Although single photons are themselves probabilistic non-Gaussian resources, they are fundamentally much more accessible compared to the simultaneous generation of multiple complex non-Gaussian states~\cite{etesse2014proposal,eaton2019non,esakk2024distill}. 
    
    In this paper, we propose a novel protocol to generate hybrid entangled states. Our protocol relies solely on linear optical elements and homodyne measurements and thus experimentally feasible and resource-efficient.
    Extending the concept of small-amplitude cat states as a primary resource, this method provides a practical route toward hybrid quantum error correction. 
    We incorporate a cat breeding stage prior to the main interference circuit to achieve a greater departure from Gaussianity. We interfere multiple cat states with small amplitudes at beam splitters and perform conditional homodyne detection. This breeding process progressively increases the departure from Gaussianity for the resource states.
    Using these bred states as inputs, our protocol generates hybrid entangled states while significantly relaxing the stringent requirements for the initial resource states.
    Furthermore, the protocol admits a direct generalization of hybrid entanglement from qubit to qudit encodings, which we explicitly demonstrate for qutrits.

    This paper is organized as follows. In Sec. II, we introduce the optical resource states and describe the proposed scheme, together with its fidelity analysis. In Sec. III, we discuss enhanced protocols based on cat breeding and present a proof-of-principle extension to a hybrid qutrit state. We note that the present analysis does not include experimental noise sources such as photon loss and detection inefficiency. Finally, Sec. IV concludes the paper. 
    \begin{figure}[t] 
        \centering  \includegraphics[width=0.45\textwidth]{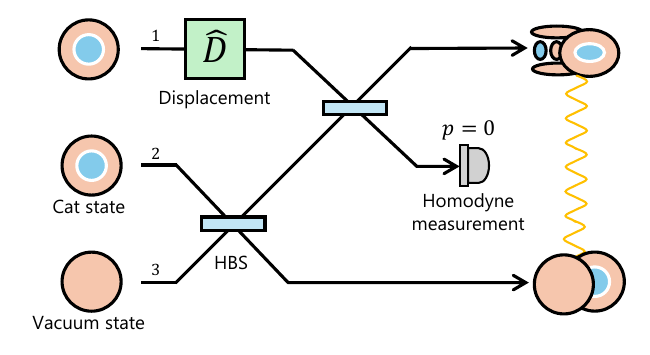} 
        \caption{Schematic of the proposed protocol. Inputs are initialized as cat states and a vacuum state. The circuit employs beam splitters, a displacement operation and a homodyne measurement ($p=0$) to generate the target hybrid entangled state.} \label{PaperFIG1_opticalSetup}
    \end{figure} 
    
\section{PRELIMINARIES AND GENERATION PROTOCOL}\label{sec1}
    In this section, we establish the theoretical framework for the proposed scheme. We first define the fundamental optical states employed as resources and then detail the protocol for generating hybrid entangled states using a breeding process based on linear optical elements.
    
    \subsection{Optical State Definitions}
    We consider the infinite-dimensional Hilbert space of a quantum harmonic oscillator. The coherent state $\ket{\alpha}$, defined as an eigenstate of the annihilation operator $\hat{a}$, is expressed in the Fock basis $\ket{n}$ as
    \begin{equation*}
        \ket{\alpha} = \hat{D}(\alpha)\ket{0} = e^{\alpha\hat{a}^\dagger-\alpha^* \hat{a}}\ket{0} = e ^{-\frac{|\alpha|^2}{2}}\sum_{n=0}^{\infty}\frac{\alpha ^n}{\sqrt{n!}}\ket{n},
    \end{equation*}
    where $\hat{D}(\alpha) = \exp(\alpha\hat{a}^\dagger - \alpha^*\hat{a})$ is the displacement operator.
    To encode quantum information, we utilize Schrödinger's cat states, which are superpositions of coherent states. The even and odd cat states are defined as
    \begin{equation}\label{eq:BetaCat}
        \ket{\mathcal{C}^\pm_{\alpha}}=\frac{1}{\mathcal{N}^\pm_{\alpha}}(\ket{\alpha} \pm \ket{-\alpha}) \ ,
    \end{equation}
    where the normalization constants $\mathcal{N}_{\alpha}^{\pm}$ are given by $\mathcal{N}_{\alpha}^{\pm} = \sqrt{2(1\pm e^{-2|\alpha|^2})}$.
    In the regime of small amplitude ($\alpha \ll 1$), the odd cat state $\ket{\mathcal{C}_{\alpha}^{-}}$ serves as an excellent approximation of a single-photon state. For small $\alpha$, this state approximates to
    \begin{equation}\label{eq:Cat_approximation}
        \ket{\mathcal{C}^-_{\alpha}} \approx \frac{e^{-|\alpha|^2/2}}{\mathcal{N}_{\alpha}^{-}} (2\alpha\ket{1} + \mathcal{O}(\alpha^3)) \sim \ket{1}.
    \end{equation}
    The fidelity between the odd cat state and the ideal single-photon state is given by 
    \begin{equation}\label{eq:Cat}
        F = |\braket{1}{\mathcal{C}^-_{\alpha}}|^2 = \frac{4|\alpha|^2 e^{-|\alpha|^2}}{(\mathcal{N}^-_{\alpha})^2},
    \end{equation}
    which approaches unity as $\alpha \to 0$. 

    \begin{figure}[t] 
        \centering  \includegraphics[width=0.45\textwidth]{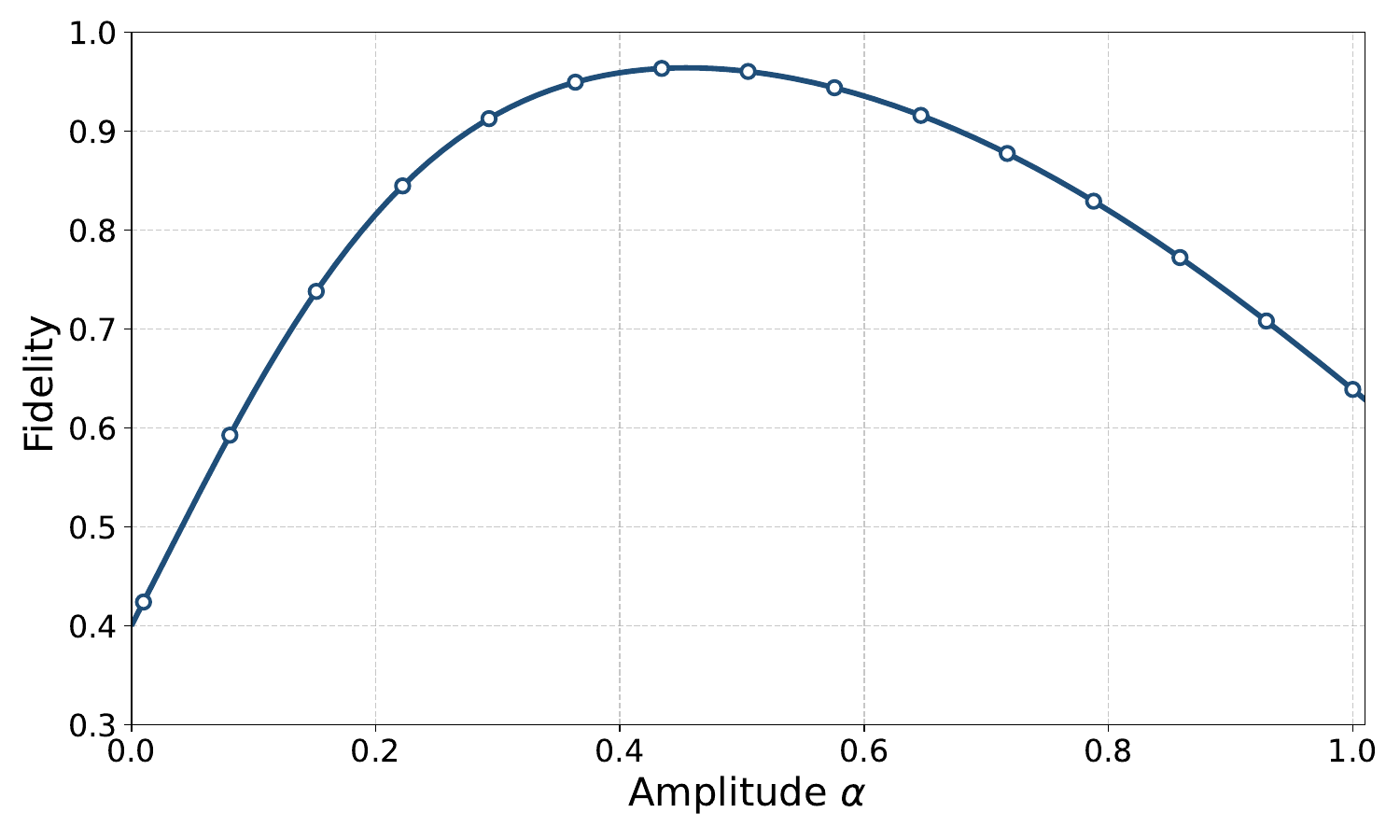} 
        \caption{Fidelity between the ideal target state $\ket{\psi_o}$ and the generated state $\ket{\psi'}$. The optimal value appears at a finite amplitude, $\alpha \approx 0.455$, resulting from trade-off between the small-amplitude approximation and the overlap with the target state, rather than $\alpha \to 0$.
    } \label{PaperFIG_APPENDIX_Fidelity}
    \end{figure} 

    \subsection{Protocol for Hybrid State Generation}
    We describe the core protocol for generating hybrid entangled states. The proposed optical setup consists of three spatial modes, as illustrated in Fig.~\ref{PaperFIG1_opticalSetup}.
    The initial state $|\psi_{i}\rangle$ is prepared as a product of two odd cat states in modes 1 and 2, and a vacuum state in mode 3:
    \begin{equation}\label{eq:Theory_initial}
        \ket{\psi_{\text{i}}} = \ket{\mathcal{C}^-_{\alpha}}_1 \otimes \ket{\mathcal{C}^-_{\sqrt{2}\alpha}}_2 \otimes \ket{0}_3.
    \end{equation}
    First, we apply a half beam splitter (HBS) operation $\hat{B}_{23}$ between modes 2 and 3. In the limit of small amplitude, this transforms the state into an entangled state:
    \begin{equation}\label{eq:approximate_trans}
    \begin{split}
        \hat{B}_{23}\ket{\psi_{\text{i}}} &= \ket{\mathcal{C}^-_{\alpha}}_1 \otimes \frac{1}{\sqrt{2}}\left( \ket{\tilde{\mathcal{C}}^-_\alpha}_2 \ket{\tilde{\mathcal{C}}^+_\alpha}_3 + \ket{\tilde{\mathcal{C}}^+_\alpha}_2 \ket{\tilde{\mathcal{C}}^-_\alpha}_3 \right) \\
        &\approx \ket{\mathcal{C}^-_{\alpha}}_1 \otimes \frac{1}{\sqrt{2}}\left( \ket{\mathcal{C}^-_{\alpha}}_2\ket{0}_3 + \ket{0}_2\ket{1}_3 \right),
    \end{split}
    \end{equation}
    where we have defined $\ket{\tilde{\mathcal{C}}^\pm_{\alpha}}=\ket{\alpha} \pm \ket{-\alpha}$.
    This approximation is discussed in Appendix~\ref{Sec:Approximation}.
    
    Next, a displacement operation $\hat{D}_{1}(\alpha)$ is applied to mode 1. This operation displaces the amplitude of the mode-1 component and prepares it for subsequent interference.

    Finally, modes 1 and 2 are interfered using a second HBS $\hat{B}_{12}$, followed by a conditional measurement on mode 2. By post-selecting the outcome where a specific quadrature is measured ($p=0$), we project the system into the target state.
    For initial states composed of superpositions of coherent states $\ket{\pm \alpha}$, the pre-measurement state can be obtained analytically. 
    The state in the output modes 1 and 3 reads
    \begin{equation}\label{Eq:OutStateBeta}\begin{split}
        \ket{\psi_o} &\propto 
        \left(\ket{3\beta}_1 - 2\ket{\beta}_1 + \ket{-\beta}_1 \right)\ket{0}_3 \\
        &+ \left(\ket{2\beta}_1 - \ket{0}_1\right)\ket{1}_3 ,
    \end{split}\end{equation}
    where $\beta \equiv \alpha/\sqrt{2}$. 
    The state in mode 1 represents an approximate GKP state~\cite{Gottesman2001GKP,Konno2024GKP}. The logical qubit is defined by
    \begin{subequations}
        \begin{equation}
            \ket{\tilde{0}_L} \propto \ket{3\beta} - 2\ket{\beta} + \ket{-\beta},
        \end{equation}
        \begin{equation}
            \ket{\tilde{1}_L} \propto \ket{2\beta} - \ket{0} .
        \end{equation}
    \end{subequations}
    This result indicates that the output state exhibits entanglement between the logical qubit encoded in mode 1 and the photon-number states in mode 3. 
    \begin{figure}[t] 
        \centering  \includegraphics[width=0.45\textwidth]{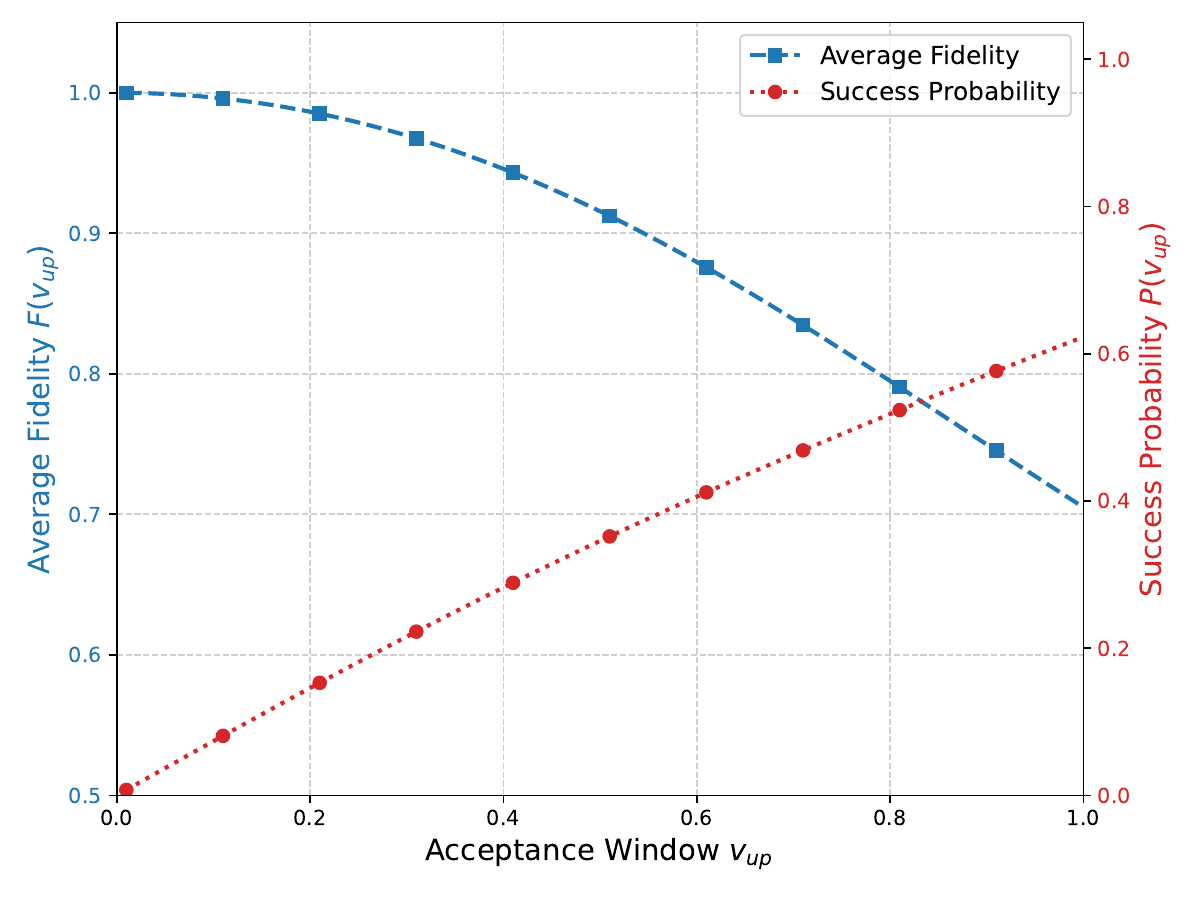} 
        \caption{We plot the average fidelity and success probability against the homodyne acceptance window $v_{up}$. This reveals a clear trade-off between the two metrics. The dashed curve indicates the average fidelity $F(v_{up})$. The dotted curve traces the success probability $P(v_{up})$. The resource state amplitude is fixed at the optimal value of $\alpha \approx 0.455$.} \label{PaperFIG3_AverageFidelity}
    \end{figure} 
\subsection{Fidelity and Limits of the Small-Amplitude Approximation}
     Our protocol is based on a single-photon approximation given by Eq.(\ref{eq:Cat}). We analyze the infidelity introduced by this assumption. For a finite amplitude $\alpha$, the exact output state without the approximation takes the form
     \begin{equation}\label{ApEq:NonAppOutput}\begin{split}
        \ket{\psi'}_{13} &\propto \left( \ket{\frac{3\alpha}{\sqrt{2}}}_1 - \ket{\frac{\alpha}{\sqrt{2}}}_1 \right)\ket{\alpha}_3 \\
        &- \left( \ket{\frac{\alpha}{\sqrt{2}}}_1 - \ket{-\frac{\alpha}{\sqrt{2}}}_1 \right) \ket{-\alpha}_3.
    \end{split}\end{equation}
    We performed numerical simulations to calculate the fidelity $F$ between this exact output state and the approximate target state defined in Eq.(\ref{Eq:OutStateBeta}). 
    
    The simulation results are plotted in Fig.~\ref{PaperFIG_APPENDIX_Fidelity}.
    Numerical simulations show that the fidelity attains a maximum value of $F \approx 0.964$ at an optimal amplitude of $\alpha \approx 0.455$.
    In the limit $\alpha \to 0$, the fidelity approaches 0.4. This behavior can be attributed to the asymptotic structure of the states in the small-amplitude regime. As $\alpha$ decreases, the target state $|\psi_o\rangle$ becomes dominated by the single-photon component $|1\rangle_3$. In contrast, the generated state $|\psi'\rangle$ retains a significant vacuum contribution $|0\rangle_3$ of comparable order. This discrepancy in the state composition limits the achievable fidelity. The detailed derivation is provided in Appendix~\ref{Sec:DetailedAnalysis}.
    \begin{figure*}[t] 
        \centering  \includegraphics[width=0.95\textwidth]{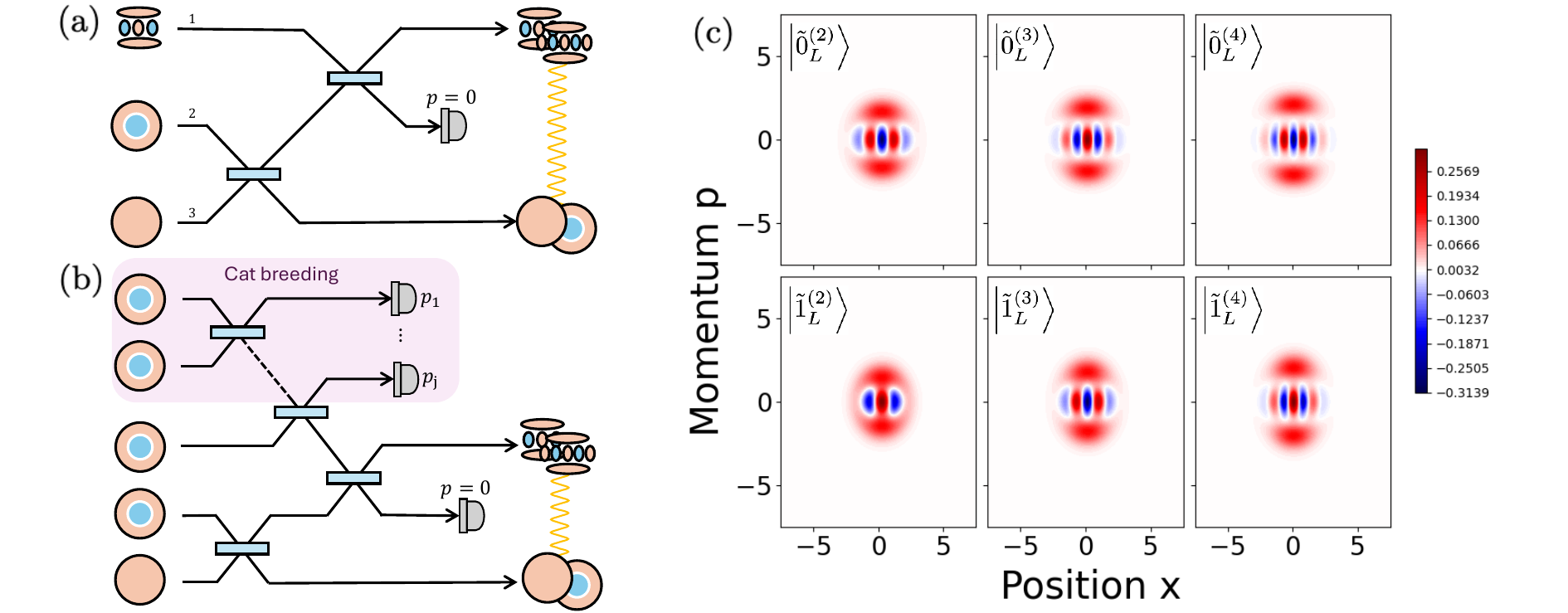} 
        \caption{Enhanced protocol via cat breeding.
        (a) Optical setup. We use a high non-Gaussianity state $\ket{\tilde{0}_{L}^{(j)}}$ as the input state instead of a single photon.
        (b) Equivalent circuit using the breeding protocol with all input states as single photons.
        (c) Wigner function of the output hybrid state. It demonstrates the enhanced non-Gaussianity achieved by breeding.} \label{PaperFIG2_CatBreeding}
    \end{figure*}
\subsection{Average Fidelity}
    We introduce $\ket{g(p)}$ as the generated unnormalized state for a probability amplitude at $p$. This state takes the form
    \begin{equation}
        \ket{g(p)} \propto \big( \ket{\phi_0(p)}\ket{0}_3 + \ket{\phi_1(p)}\ket{1}_3 \big).
    \end{equation}
    The detailed derivation of $\ket{g(p)}$ is provided in Appendix~\ref{HomodyneError}.
    We then denote the normalized target state by $\ket{t}$. This target state corresponds to the ideal case at $p=0$ and is given by
    \begin{equation}
        \ket{t} \propto \ket{g(0)}= \ket{\psi_o}.
    \end{equation}
    A measurement is deemed successful if the outcome falls within $|p| < v_{up}$. The probability of these events is determined by integrating the state norm. The result is given by
    \begin{equation}
    P(v_{up}) = \frac{1}{P_{all}} \int_{-v_{up}}^{v_{up}} \braket{g(p)}{g(p)} dp
    \end{equation}
    An infinite integration range defines the total probability $P_{all}$. A probability-weighted average of the state fidelity $F'(p)$ between the generated state at a specific outcome $p$ and the ideal target state $\ket{t}$ over the acceptance window then leads to the final expression. We write this result as
    \begin{equation}\begin{split}
    F(v_{up}) &= \frac{\int_{-v_{up}}^{v_{up}} F'(p) \braket{g(p)}{g(p)} dp}{\int_{-v_{up}}^{v_{up}} \braket{g(p)}{g(p)} dp} \\
    &= \frac{\int_{-v_{up}}^{v_{up}} |\braket{t}{g(p)}|^2 dp}{\int_{-v_{up}}^{v_{up}} \braket{g(p)}{g(p)} dp}
    \end{split}\end{equation}
    
    We assess the experimental feasibility of our protocol. We analyze the trade-off between average fidelity and success probability. This evaluation assumes finite homodyne acceptance windows $|p| \le v_{up}$. Fig.~\ref{PaperFIG3_AverageFidelity} illustrates these results. A moderate expansion of the acceptance window significantly increases the success probability.
    The corresponding fidelity degradation remains minimal. 
    A stringent average fidelity of 0.99 yields a practical success probability of approximately 12.6\%. An alternative configuration secures a 10\% success probability. This setup maintains an excellent average fidelity of 0.994.
    Furthermore, we establish $v_{up} \approx 0.5$ as a practical example of a relaxed window. The protocol achieves an average fidelity of approximately 0.90 in this case. The success probability simultaneously approaches a remarkably high 40\%.
    
\section{ENHANCED PROTOCOLS VIA CAT BREEDING}\label{sec2}
    \subsection{Cat Breeding Concept}\label{Sec:CatBreedingConcept}
    In the previous section, the input state in mode 1 was approximated as a single-photon state, which corresponds to an odd cat state in the limit of small amplitude ($\alpha \ll 1$). 
    This approximation alleviates experimental demands yet still requires resources with stronger non-Gaussianity. We therefore introduce a cat-breeding stage prior to the main interference circuit \cite{Weigand2018Breed,vasconcelos2010all,takase2024generation,solodovnikova2025loss}.

    It is well established that the application of iterative breeding operations to single-photon sources enables the generation of quantum states with enhanced non-Gaussianity.
    In our proposed scheme, we implement a breeding protocol in which multiple small-amplitude cat states interfere at beam splitters. 
    Subsequent homodyne detection leads to an enhancement of non-Gaussianity.
    Let $\ket{\tilde{0}_L^{(j)}}$ denote the state obtained after applying the breeding operation $j$ times to the initial small-amplitude cat states. We replace the small-amplitude cat state input in mode 1 with this bred state, such that the modified initial state of the system is given by
    \begin{equation}\label{eq:Theory_initial3}
        \ket{\psi_{\text{i}}^{(j)}} = \ket{\tilde{0}_L^{(j)}} \otimes \ket{\mathcal{C}^-_{\sqrt{2}\alpha}}_2 \otimes \ket{0}_3,
    \end{equation}
    where mode 2 is prepared in an approximate odd cat state and mode 3 is initialized in the vacuum state.

   The breeding process progressively transforms the input into a state exhibiting pronounced non-Gaussianity and approaching a GKP state as the number of operations $j$ increases.
   This enhancement in the input resource is crucial for generating hybrid entangled states with sufficient quality for logical encoding. As illustrated in the optical setup shown in Fig.~\ref{PaperFIG2_CatBreeding}(a), the output state exhibits entanglement between a logical GKP qubit and a single-photon state, enabled by the enhanced non-Gaussianity generated in the breeding stage.

    We consider the initial state $\ket{\tilde{0}_L}$.
    We first apply the HBS $\hat{B}_{23}$ to the total input state, yielding
    \begin{equation}
        \hat{B}_{23}\ket{\psi_{\text{i}}} \approx  \ket{\tilde{0}_L}_1 \otimes \frac{1}{\sqrt{2}}\left( \ket{1}_2\ket{0}_3 + \ket{0}_2\ket{1}_3 \right),
    \end{equation}
    where the single-photon approximation in Eq. (\ref{eq:Cat_approximation}) has been employed.
    Subsequently, we apply an HBS operation $\hat{B}_{12}$ between modes 1 and 2. 
    A homodyne measurement is then performed on mode 2. Conditioned on the outcome $p=0$, it yields the output state:
    \begin{equation}\label{Eq:OutStateBeta2}\begin{split}
        \ket{\psi_o^{(2)}} &\propto 
        \ket{\tilde{0}^{(2)}_L}_1\ket{0}_3 + \ket{\tilde{1}^{(2)}_L}_1\ket{1}_3,
    \end{split}\end{equation}
    where 
    \begin{subequations}
        \begin{equation}\begin{split}
            \ket{\tilde{0}^{(2)}_L} &= \ket{\frac{3\alpha}{2} + \frac{\alpha}{\sqrt{2}}} - \ket{\frac{3\alpha}{2} - \frac{\alpha}{\sqrt{2}}}- 2\ket{\frac{\alpha}{2} + \frac{\alpha}{\sqrt{2}}} \\
            &+ 2\ket{\frac{\alpha}{2} - \frac{\alpha}{\sqrt{2}}}+ \ket{-\frac{\alpha}{2} + \frac{\alpha}{\sqrt{2}}} - \ket{-\frac{\alpha}{2} - \frac{\alpha}{\sqrt{2}}},
        \end{split}\end{equation}
        \begin{equation}
            \ket{\tilde{1}^{(2)}_L} = \ket{\frac{3\alpha}{2}} - 2\ket{\frac{\alpha}{2}} + \ket{-\frac{\alpha}{2}}.
        \end{equation}
    \end{subequations}
    This derivation demonstrates that using the bred state as an input resource yields a hybrid entangled state with pronounced non-Gaussianity. Fig.~\ref{PaperFIG2_CatBreeding}(c) shows the Wigner function of the generated state obtained from numerical calculations.
    \begin{figure}[t] 
        \centering  \includegraphics[width=0.45\textwidth]{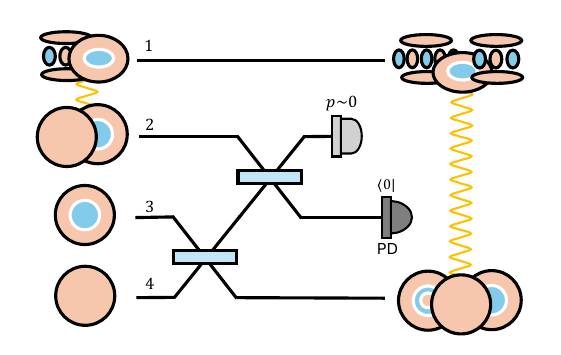} 
        \caption{Optical setup for generating a hybrid qutrit state. The input is the hybrid entangled state given by Eq. (\ref{Eq:OutStateBeta}). The hybrid qutrit state is generated when the homodyne measurement result is $p=0$ and the photon detector (PD) projects onto $_2\bra{0}$.} \label{PaperFIG3_qutrit}
    \end{figure} 
    
    \subsection{Approximate hybrid qudit generation}\label{Sec:HybridQuditGeneration}
    The GKP qubit encodes a two-level logical system in a continuous-variable state. The GKP qudit generalizes this construction to a $d$-dimensional discrete logical space \cite{schmidt2022quantum,schmidt2024error,brock2025quantum}. 
    Here, we demonstrate that entanglement between a GKP qudit and a photon-number state can also be generated using a similar optical setup to that shown in Fig.~\ref{PaperFIG2_CatBreeding}(a).
    
    We consider an initial state consisting of an approximate GKP state entangled between modes 1 and 2, and a single-photon path-entangled state in modes 3 and 4, given by
    \begin{equation}\begin{aligned}
        &\ket{\psi_o}_{12} \otimes \frac{1}{\sqrt{2}}\left( \ket{1}_3\ket{0}_4 + \ket{0}_3\ket{1}_4 \right) \\
        &  \propto \left( \ket{\tilde{0}_L}_1\ket{0}_2 + \ket{\tilde{1}_L}_1\ket{1}_2 \right)\\
         & \quad \otimes \left[ (\ket{\alpha}_3 - \ket{-\alpha}_3)\otimes\ket{0}_4 + \ket{0}_3\ket{1}_4 \right].
    \end{aligned}\end{equation}
    The corresponding optical setup is illustrated in Fig. \ref{PaperFIG3_qutrit}.
    A HBS is applied to modes 1 and 3, followed by a homodyne measurement on mode 3, yielding the post-measurement state: 
    \begin{equation}\begin{aligned}
        \ket{\psi_\textrm{abp}} \propto 
        &\frac{1}{\sqrt{2}}\bigg( 
        \ket{\tilde{0}^{(2)}_L}_1\ket{0}_2\ket{0}_4 + 
        \ket{\tilde{1}^{(2)}_L}_1\ket{0}_2\ket{1}_4 \\
        &+ \ket{1_\textrm{qt}}_1\ket{1}_2\ket{0}_4 +
        \ket{2_\textrm{qt}}_1\ket{1}_2\ket{1}_4 \bigg).
    \end{aligned}\end{equation}
    Here, the unnormalized conditional states in mode 1 are given by 
    \begin{subequations}
        \begin{equation}
            \ket{1_\textrm{qt}} \approx \ket{\alpha + \frac{\alpha}{\sqrt{2}}} - \ket{\alpha - \frac{\alpha}{\sqrt{2}}} - \ket{\frac{\alpha}{\sqrt{2}}} + \ket{-\frac{\alpha}{\sqrt{2}}},
        \end{equation}
        \begin{equation}
            \ket{2_\textrm{qt}}=\ket{\alpha} - \ket{0}.
        \end{equation}
    \end{subequations}
    Subsequently, a HBS is applied to modes 2 and 4. Finally, projecting mode 2 onto the vacuum state $\ket{0}_2$ results in the hybrid qudit–photon state:
    \begin{equation}
        \ket{\psi_\textrm{hqt}}\propto  
        \ket{\tilde{0}^{(2)}_L}_1\ket{0}_4 + 
        \big(\ket{\tilde{1}^{(2)}_L}_1 + 
        \ket{1_\textrm{qt}}_1\big)\ket{1}_4 +
        \ket{2_\textrm{qt}}_1\ket{2}_4 \quad.
    \end{equation}
    The state $\ket{\tilde{0}^{(2)}_L}$ is orthogonal to $\ket{\tilde{1}^{(2)}_L} + \ket{1_\textrm{qt}}$, and $\ket{\tilde{1}^{(2)}_L} + \ket{1_\textrm{qt}}$ is orthogonal to $\ket{2_\textrm{qt}}$. 
    By contrast, $\ket{\tilde{0}^{(2)}_L}$ and $\ket{2_\mathrm{qt}}$ are not strictly orthogonal owing to a small residual overlap.
    This overlap can be suppressed by increasing the amplitude $\alpha$ or by applying additional squeezing, thereby approximately restoring orthogonality.
    A scheme for generating ideal GKP qutrit states is presented in Appendix \ref{App:qutrit}. We presented the qutrit extension only as a proof-of-principle example to illustrate the extensibility of our setup. Since the main focus of the present work is the generation of hybrid GKP qubit entanglement, we leave a more systematic analysis for future work.
    
\section{CONCLUSION}
    In this paper, we proposed a novel scheme to generate hybrid GKP entangled states using only small-amplitude cat states as non-Gaussian input resources.
    Our method utilizes a small-amplitude cat to create entanglement between a GKP qubit and a photon-number state.
    Our approach requires only linear optics and homodyne detection, making it experimentally accessible with current optical technology \cite{madsen2022quantum,maring2024versatile,tomoda2024boosting,Konno2024GKP,hanamura2025scalable,aghaee2025scaling,larsen2025integrated,kala2025nullifiers,breum2025distribution}.

    We demonstrated that the breeding process using small-amplitude cats enhances the non-Gaussianity of the input state. 
    Our scheme is resource-efficient compared to conventional methods that require multiple probabilistic non-Gaussian resources. Consequently, it has the potential to overcome the bottleneck in generation rates that typically arise in hybrid state generation protocols.

    Furthermore, we demonstrated the extensibility of our protocol. Specifically, it can be applied to generate hybrid GKP qudit states, as explicitly demonstrated through a concrete construction of a hybrid qutrit state.
    This work contributes to lowering the hurdles for fault-tolerant quantum computation, such as the requirement for high squeezing levels. We note that this work does not explicitly include experimental imperfections, and incorporating effects such as photon loss is left for future work. Further extensions to other hybrid bosonic codes, including qutrit generation, are also left for future work.

    As a consequence, the hybrid states produced by the proposed method are highly promising for cost-effective and efficient quantum computation. Furthermore, their utility extends directly to quantum communication \cite{fukui2024resource}. In particular, they are highly suitable for advanced protocols, including quantum repeaters \cite{fukui2021all,schmidt2022quantum,schmidt2024error,haussler2025long,haussler2025quantum,chelluri2025bosonic,haussler2025quantum2} and entanglement sharing \cite{bose2024long}.
    
\section*{Acknowledgements} 
    We thank Takaya Matsuura, Shuntaro Takeda, Akihiro Machinaga and Ryoga Sakurada for helpful discussions. 
    This work was supported by the Advancement of Next Generation Research Projects, Keio University; JST Moonshot R\&D Grant Nos.~JPMJMS2061 and JPMJMS2064; JST ASPIRE Grant No.~JPMJAP2427; JST SPRING Grant No.~JPMJSP2123; JST PRESTO Grant No.~JPMJPR23FA; JSPS KAKENHI Grant No.~25K22795; UTokyo Foundation; and donations from Nichia Corporation.
    %This work was supported by the Advancement of Next Generation Research Projects, Keio University; JST Moonshot R\&D, JPMJMS2061; JST SPRING, JPMJSP2123. KF acknowledges support from JST PRESTO Grant No.~JPMJPR23FA, JSPS KAKENHI Grant No.~25K22795, JST Moonshot R\&D Grant No.~JPMJMS2064, JST Moonshot R\&D Grant No.~JPMJMS2061, UTokyo Foundation, and donations from Nichia Corporation.
    
\section*{DATA AVAILABILITY} 
    The data that support the findings of this article are not publicly available. The data are available from the authors upon reasonable request.

\bibliography{main} %hoge.bibから拡張子を外した名前

@article{shor1995scheme,
  title={Scheme for reducing decoherence in quantum computer memory},
  author={Shor, Peter W},
  journal={Physical review A},
  volume={52},
  number={4},
  pages={R2493},
  year={1995},
  publisher={APS}
}

@article{kitaev2003fault,
  title={Fault-tolerant quantum computation by anyons},
  author={Kitaev, A Yu},
  journal={Annals of physics},
  volume={303},
  number={1},
  pages={2--30},
  year={2003},
  publisher={Elsevier}
}

@article{horsman2012surface,
  title={Surface code quantum computing by lattice surgery},
  author={Horsman, Dominic and Fowler, Austin G and Devitt, Simon and Van Meter, Rodney},
  journal={New Journal of Physics},
  volume={14},
  number={12},
  pages={123011},
  year={2012},
  publisher={IOP Publishing}
}

@article{Gottesman2001GKP,
  title = {Encoding a qubit in an oscillator},
  author = {Gottesman, Daniel and Kitaev, Alexei and Preskill, John},
  journal = {Phys. Rev. A},
  volume = {64},
  issue = {1},
  pages = {012310},
  numpages = {21},
  year = {2001},
  publisher = {American Physical Society},
  doi = {10.1103/PhysRevA.64.012310},
  url = {https://link.aps.org/doi/10.1103/PhysRevA.64.012310}
}

@article{Fukui2017GkpAnalog,
  title = {Analog Quantum Error Correction with Encoding a Qubit into an Oscillator},
  author = {Fukui, Kosuke and Tomita, Akihisa and Okamoto, Atsushi},
  journal = {Phys. Rev. Lett.},
  volume = {119},
  issue = {18},
  pages = {180507},
  numpages = {4},
  year = {2017},
  publisher = {American Physical Society},
  doi = {10.1103/PhysRevLett.119.180507},
  url = {https://link.aps.org/doi/10.1103/PhysRevLett.119.180507}
}

@article{Albert2018GkpLoss,
  title = {Performance and structure of single-mode bosonic codes},
  author = {Albert, Victor V. and Noh, Kyungjoo and Duivenvoorden, Kasper and Young, Dylan J. and Brierley, R. T. and Reinhold, Philip and Vuillot, Christophe and Li, Linshu and Shen, Chao and Girvin, S. M. and Terhal, Barbara M. and Jiang, Liang},
  journal = {Phys. Rev. A},
  volume = {97},
  issue = {3},
  pages = {032346},
  numpages = {30},
  year = {2018},
  publisher = {American Physical Society},
  doi = {10.1103/PhysRevA.97.032346},
  url = {https://link.aps.org/doi/10.1103/PhysRevA.97.032346}
}

@article{Weigand2018Breed,
  title = {Generating grid states from Schr\"odinger-cat states without postselection},
  author = {Weigand, Daniel J. and Terhal, Barbara M.},
  journal = {Phys. Rev. A},
  volume = {97},
  issue = {2},
  pages = {022341},
  numpages = {13},
  year = {2018},
  publisher = {American Physical Society},
  doi = {10.1103/PhysRevA.97.022341},
  url = {https://link.aps.org/doi/10.1103/PhysRevA.97.022341}
}

@article{Konno2024GKP,
    author = {Shunya Konno  and Warit Asavanant  and Fumiya Hanamura  and Hironari Nagayoshi  and Kosuke Fukui  and Atsushi Sakaguchi  and Ryuhoh Ide  and Fumihiro China  and Masahiro Yabuno  and Shigehito Miki  and Hirotaka Terai  and Kan Takase  and Mamoru Endo  and Petr Marek  and Radim Filip  and Peter van Loock  and Akira Furusawa },
    title = {Logical states for fault-tolerant quantum computation with propagating light},
    journal = {Science},
    volume = {383},
    number = {6680},
    pages = {289-293},
    year = {2024},
    doi = {10.1126/science.adk7560},
    URL = {https://www.science.org/doi/abs/10.1126/science.adk7560},
    eprint = {https://www.science.org/doi/pdf/10.1126/science.adk7560},
}

@article{larsen2025integrated,
  title={Integrated photonic source of Gottesman--Kitaev--Preskill qubits},
  author={Larsen, MV and Bourassa, JE and Kocsis, S and Tasker, JF and Chadwick, RS and Gonz{\'a}lez-Arciniegas, C and Hastrup, J and Lopetegui-Gonz{\'a}lez, CE and Miatto, FM and Motamedi, A and others},
  journal={Nature},
  pages={1--5},
  year={2025},
  publisher={Nature Publishing Group UK London}
}

@article{ralph2003quantum,
  title={Quantum computation with optical coherent states},
  author={Ralph, Timothy C and Gilchrist, Alexei and Milburn, Gerard J and Munro, William J and Glancy, Scott},
  journal={Physical Review A},
  volume={68},
  number={4},
  pages={042319},
  year={2003},
  publisher={APS}
}

@article{michael2016new,
  title={New class of quantum error-correcting codes for a bosonic mode},
  author={Michael, Marios H and Silveri, Matti and Brierley, RT and Albert, Victor V and Salmilehto, Juha and Jiang, Liang and Girvin, Steven M},
  journal={Physical Review X},
  volume={6},
  number={3},
  pages={031006},
  year={2016},
  publisher={APS}
}

@article{su2019conversion,
  title={Conversion of Gaussian states to non-Gaussian states using photon-number-resolving detectors},
  author={Su, Daiqin and Myers, Casey R and Sabapathy, Krishna Kumar},
  journal={Physical Review A},
  volume={100},
  number={5},
  pages={052301},
  year={2019},
  publisher={APS}
}

@article{bourassa2021blueprint,
  title={Blueprint for a scalable photonic fault-tolerant quantum computer},
  author={Bourassa, J. Eli and Alexander, Rafael N and Vasmer, Michael and Patil, Ashlesha and Tzitrin, Ilan and Matsuura, Takaya and Su, Daiqin and Baragiola, Ben Q and Guha, Saikat and Dauphinais, Guillaume and others},
  journal={Quantum},
  volume={5},
  pages={392},
  year={2021},
  publisher={Verein zur F{\"o}rderung des Open Access Publizierens in den Quantenwissenschaften}
}

@article{su2022universal,
  title={Universal quantum computation with optical four-component cat qubits},
  author={Su, Daiqin and Dhand, Ish and Ralph, Timothy C},
  journal={Physical Review A},
  volume={106},
  number={4},
  pages={042614},
  year={2022},
  publisher={APS}
}

@article{takase2023gottesman,
  title={Gottesman-Kitaev-Preskill qubit synthesizer for propagating light},
  author={Takase, Kan and Fukui, Kosuke and Kawasaki, Akito and Asavanant, Warit and Endo, Mamoru and Yoshikawa, Jun-ichi and van Loock, Peter and Furusawa, Akira},
  journal={npj Quantum Information},
  volume={9},
  number={1},
  pages={98},
  year={2023},
  publisher={Nature Publishing Group UK London}
}

@article{endo2025high,
  title={High-Rate Four Photon Subtraction from Squeezed Vacuum: Preparing Cat State for Optical Quantum Computation},
  author={Endo, Mamoru and Nomura, Takefumi and Sonoyama, Tatsuki and Takahashi, Kazuma and Takasu, Sachiko and Fukuda, Daiji and Kashiwazaki, Takahiro and Inoue, Asuka and Umeki, Takeshi and Nehra, Rajveer and others},
  journal={arXiv preprint arXiv:2502.08952},
  year={2025}
}

@article{kwon2014generation,
  title = {Generation of hybrid entanglement between a single-photon polarization qubit and a coherent state},
  author = {Kwon, Hyukjoon and Jeong, Hyunseok},
  journal = {Phys. Rev. A},
  volume = {91},
  issue = {1},
  pages = {012340},
  numpages = {6},
  year={2015},
  publisher = {American Physical Society},
  doi = {10.1103/PhysRevA.91.012340},
  url = {https://link.aps.org/doi/10.1103/PhysRevA.91.012340}
}

@article{andersen2015hybrid,
  title={Hybrid discrete-and continuous-variable quantum information},
  author={Andersen, Ulrik L and Neergaard-Nielsen, Jonas S and Van Loock, Peter and Furusawa, Akira},
  journal={Nature Physics},
  volume={11},
  number={9},
  pages={713--719},
  year={2015},
  publisher={Nature Publishing Group UK London}
}

@article{lee2013near,
  title={Near-deterministic quantum teleportation and resource-efficient quantum computation using linear optics and hybrid qubits},
  author={Lee, Seung-Woo and Jeong, Hyunseok},
  journal={Physical Review A—Atomic, Molecular, and Optical Physics},
  volume={87},
  number={2},
  pages={022326},
  year={2013},
  publisher={APS}
}

@article{jeong2014generation,
  title={Generation of hybrid entanglement of light},
  author={Jeong, Hyunseok and Zavatta, Alessandro and Kang, Minsu and Lee, Seung-Woo and Costanzo, Luca S and Grandi, Samuele and Ralph, Timothy C and Bellini, Marco},
  journal={Nature Photonics},
  volume={8},
  number={7},
  pages={564--569},
  year={2014},
  publisher={Nature Publishing Group UK London}
}

@article{lee2015nearly,
  title={Nearly deterministic Bell measurement for multiphoton qubits and its application to quantum information processing},
  author={Lee, Seung-Woo and Park, Kimin and Ralph, Timothy C and Jeong, Hyunseok},
  journal={Physical review letters},
  volume={114},
  number={11},
  pages={113603},
  year={2015},
  publisher={APS}
}

@article{omkar2020resource,
  title={Resource-efficient topological fault-tolerant quantum computation with hybrid entanglement of light},
  author={Omkar, Srikrishna and Teo, Yong Siah and Jeong, Hyunseok},
  journal={Physical Review Letters},
  volume={125},
  number={6},
  pages={060501},
  year={2020},
  publisher={APS}
}

@article{omkar2021highly,
  title={Highly photon-loss-tolerant quantum computing using hybrid qubits},
  author={Omkar, Srikrishna and Teo, YS and Lee, Seung-Woo and Jeong, Hyunseok},
  journal={Physical Review A},
  volume={103},
  number={3},
  pages={032602},
  year={2021},
  publisher={APS}
}

@article{lee2024fault,
  title={Fault-tolerant quantum computation by hybrid qubits with bosonic cat code and single photons},
  author={Lee, Jaehak and Kang, Nuri and Lee, Seok-Hyung and Jeong, Hyunseok and Jiang, Liang and Lee, Seung-Woo},
  journal={PRX Quantum},
  volume={5},
  number={3},
  pages={030322},
  year={2024},
  publisher={APS}
}

@article{fukui2024resource,
  title={Resource-efficient high-threshold fault-tolerant quantum computation with weak nonlinear optics},
  author={Fukui, Kosuke and van Loock, Peter},
  journal={arXiv preprint arXiv:2412.16536},
  year={2024}
}

@article{kiryu2025linear,
  title={Linear optical quantum computing with a hybrid squeezed-cat code},
  author={Kiryu, Shohei and Fukui, Kosuke and Okamoto, Atsushi and Tomita, Akihisa},
  journal={Physical Review A},
  volume={112},
  number={4},
  pages={042615},
  year={2025},
  publisher={APS}
}

@article{bera2025long,
  title={Long-distance Bell nonlocality and teleportation with hybrid entangled states},
  author={Bera, Subhankar and Bose, Soumyakanti and Jeong, Hyunseok and Majumdar, Archan S},
  journal={arXiv e-prints},
  pages={arXiv--2502},
  year={2025}
}

@article{etesse2014proposal,
  title={Proposal for a loophole-free violation of Bell's inequalities with a set of single photons and homodyne measurements},
  author={Etesse, Jean and Blandino, R{\'e}mi and Kanseri, Bhaskar and Tualle-Brouri, Rosa},
  journal={New Journal of Physics},
  volume={16},
  number={5},
  pages={053001},
  year={2014},
  publisher={IOP Publishing}
}

@article{eaton2019non,
  title={Non-Gaussian and Gottesman--Kitaev--Preskill state preparation by photon catalysis},
  author={Eaton, Miller and Nehra, Rajveer and Pfister, Olivier},
  journal={New Journal of Physics},
  volume={21},
  number={11},
  pages={113034},
  year={2019},
  publisher={IOP Publishing}
}

@article{esakk2024distill,
  title={Distillation of continuous variable qudits from single photon sources: A cascaded approach},
  author={Devibala Esakkimuthu and Basherrudin Mahmud Ahmed Abduljaffer},
  journal={arXiv preprint arXiv:2512.13264},
  year={2025},
  url={https://arxiv.org/abs/2512.13264},
}

@article{haussler2025long,
  title={Long-distance quantum communication sending single photons and keeping many},
  author={H{\"a}ussler, Stefan and van Loock, Peter},
  journal={arXiv preprint arXiv:2512.18767},
  year={2025}
}

@article{vasconcelos2010all,
  title={All-optical generation of states for “Encoding a qubit in an oscillator”},
  author={Vasconcelos, Hilma M and Sanz, Liliana and Glancy, Scott},
  journal={Optics letters},
  volume={35},
  number={19},
  pages={3261--3263},
  year={2010},
  publisher={Optical Society of America}
}

@article{takase2024generation,
  title={Generation of flying logical qubits using generalized photon subtraction with adaptive Gaussian operations},
  author={Takase, Kan and Hanamura, Fumiya and Nagayoshi, Hironari and Bourassa, J Eli and Alexander, Rafael N and Kawasaki, Akito and Asavanant, Warit and Endo, Mamoru and Furusawa, Akira},
  journal={Physical Review A},
  volume={110},
  number={1},
  pages={012436},
  year={2024},
  publisher={APS}
}

@article{solodovnikova2025loss,
  title={The loss tolerance of cat breeding for fault-tolerant grid state generation},
  author={Solodovnikova, Olga and Andersen, Ulrik L and Neergaard-Nielsen, Jonas S},
  journal={arXiv preprint arXiv:2508.06193},
  year={2025}
}

@article{schmidt2022quantum,
  title={Quantum error correction with higher Gottesman-Kitaev-Preskill codes: Minimal measurements and linear optics},
  author={Schmidt, Frank and van Loock, Peter},
  journal={Physical Review A},
  volume={105},
  number={4},
  pages={042427},
  year={2022},
  publisher={APS}
}

@article{schmidt2024error,
  title={Error-corrected quantum repeaters with Gottesman-Kitaev-Preskill qudits},
  author={Schmidt, Frank and Miller, Daniel and van Loock, Peter},
  journal={Physical Review A},
  volume={109},
  number={4},
  pages={042427},
  year={2024},
  publisher={APS}
}

@article{brock2025quantum,
  title={Quantum error correction of qudits beyond break-even},
  author={Brock, Benjamin L and Singh, Shraddha and Eickbusch, Alec and Sivak, Volodymyr V and Ding, Andy Z and Frunzio, Luigi and Girvin, Steven M and Devoret, Michel H},
  journal={Nature},
  volume={641},
  number={8063},
  pages={612--618},
  year={2025},
  publisher={Nature Publishing Group UK London}
}

@article{madsen2022quantum,
  title={Quantum computational advantage with a programmable photonic processor},
  author={Madsen, Lars S and Laudenbach, Fabian and Askarani, Mohsen Falamarzi and Rortais, Fabien and Vincent, Trevor and Bulmer, Jacob FF and Miatto, Filippo M and Neuhaus, Leonhard and Helt, Lukas G and Collins, Matthew J and others},
  journal={Nature},
  volume={606},
  number={7912},
  pages={75--81},
  year={2022},
  publisher={Nature Publishing Group UK London}
}

@article{maring2024versatile,
  title={A versatile single-photon-based quantum computing platform},
  author={Maring, Nicolas and Fyrillas, Andreas and Pont, Mathias and Ivanov, Edouard and Stepanov, Petr and Margaria, Nico and Hease, William and Pishchagin, Anton and Lema{\^\i}tre, Aristide and Sagnes, Isabelle and others},
  journal={Nature Photonics},
  volume={18},
  number={6},
  pages={603--609},
  year={2024},
  publisher={Nature Publishing Group UK London}
}

@article{tomoda2024boosting,
  title={Boosting the generation rate of squeezed single-photon states by generalized photon subtraction},
  author={Tomoda, Hiroko and Machinaga, Akihiro and Takase, Kan and Harada, Jun and Kashiwazaki, Takahiro and Umeki, Takeshi and Miki, Shigehito and China, Fumihiro and Yabuno, Masahiro and Terai, Hirotaka and others},
  journal={Physical Review A},
  volume={110},
  number={3},
  pages={033717},
  year={2024},
  publisher={APS}
}

@article{hanamura2025scalable,
  title={Scalable Optical Quantum State Synthesizer with Dual-Mode Resonator Memory},
  author={Hanamura, Fumiya and Takase, Kan and Hirota, Kazuki and Nehra, Rajveer and Lang, Florian and Miki, Shigehito and Terai, Hirotaka and Yabuno, Masahiro and Kashiwazaki, Takahiro and Inoue, Asuka and others},
  journal={Prx Quantum},
  volume={6},
  number={4},
  pages={040336},
  year={2025},
  publisher={APS}
}

@article{aghaee2025scaling,
  title={Scaling and networking a modular photonic quantum computer},
  author={Aghaee Rad, H and Ainsworth, T and Alexander, RN and Altieri, B and Askarani, MF and Baby, R and Banchi, L and Baragiola, BQ and Bourassa, JE and Chadwick, RS and others},
  journal={Nature},
  volume={638},
  number={8052},
  pages={912--919},
  year={2025},
  publisher={Nature Publishing Group UK London}
}

@article{kala2025nullifiers,
  title={Nullifiers of non-Gaussian cluster states through homodyne measurement},
  author={Kala, Vojt{\v{e}}ch and Breum, Casper A and Larsen, Mikkel V and Andersen, Ulrik L and Neergaard-Nielsen, Jonas S and Filip, Radim and Marek, Petr},
  journal={arXiv preprint arXiv:2505.21066},
  year={2025}
}

@article{breum2025distribution,
  title={Distribution of non-Gaussian states in a deployed telecommunication fiber channel},
  author={Breum, Casper A and Guo, Xueshi and Larsen, Mikkel V and Miki, Shigehito and Terai, Hirotaka and Andersen, Ulrik L and Neergaard-Nielsen, Jonas S},
  journal={arXiv preprint arXiv:2509.18080},
  year={2025}
}

@article{bose2024long,
  title={Long-distance entanglement sharing using hybrid states of discrete and continuous variables},
  author={Bose, Soumyakanti and Singh, Jaskaran and Cabello, Ad{\'a}n and Jeong, Hyunseok},
  journal={Physical Review Applied},
  volume={21},
  number={6},
  pages={064013},
  year={2024},
  publisher={APS}
}

@article{fukui2021all,
  title={All-optical long-distance quantum communication with Gottesman-Kitaev-Preskill qubits},
  author={Fukui, Kosuke and Alexander, Rafael N and van Loock, Peter},
  journal={Physical Review Research},
  volume={3},
  number={3},
  pages={033118},
  year={2021},
  publisher={APS}
}

@article{haussler2025quantum,
  title={Quantum repeaters based on stationary Gottesman-Kitaev-Preskill qubits},
  author={H{\"a}ussler, Stefan and van Loock, Peter},
  journal={Physical Review A},
  volume={111},
  number={6},
  pages={062611},
  year={2025},
  publisher={APS}
}

@article{chelluri2025bosonic,
  title={Bosonic quantum error correction with microwave cavities for quantum repeaters},
  author={Chelluri, S. Siddardha and Sharma, Sanchar and Schmidt, Frank and Kusminskiy, Silvia Viola and van Loock, Peter},
  journal={arXiv preprint arXiv:2503.21569},
  year={2025}
}

@article{haussler2025quantum2,
  title={Quantum repeaters based on stationary and flying Gottesman-Kitaev-Preskill qudits},
  author={H{\"a}ussler, Stefan and van Loock, Peter},
  journal={arXiv preprint arXiv:2508.00530},
  year={2025}
}

@article{dakna1997generating,
  title={Generating Schr{\"o}dinger-cat-like states by means of conditional measurements on a beam splitter},
  author={Dakna, Mohammed and Anhut, Tiemo and Opatrn{\`y}, T and Kn{\"o}ll, Ludwig and Welsch, D-G},
  journal={Physical Review A},
  volume={55},
  number={4},
  pages={3184},
  year={1997},
  publisher={APS}
}

@article{lund2004conditional,
  title={Conditional production of superpositions of coherent states with inefficient photon detection},
  author={Lund, AP and Jeong, H and Ralph, TC and Kim, MS},
  journal={Physical Review A—Atomic, Molecular, and Optical Physics},
  volume={70},
  number={2},
  pages={020101},
  year={2004},
  publisher={APS}
}

@article{hanamura2025beyond,
  title={Beyond Stellar Rank: Control Parameters for Scalable Optical Non-Gaussian State Generation},
  author={Hanamura, Fumiya and Takase, Kan and Nagayoshi, Hironari and Ide, Ryuhoh and Asavanant, Warit and Fukui, Kosuke and Marek, Petr and Filip, Radim and Furusawa, Akira},
  journal={arXiv preprint arXiv:2509.06255},
  year={2025}
}

@article{noh2022low,
  title={Low-overhead fault-tolerant quantum error correction with the surface-GKP code},
  author={Noh, Kyungjoo and Chamberland, Christopher and Brand{\~a}o, Fernando GSL},
  journal={PRX Quantum},
  volume={3},
  number={1},
  pages={010315},
  year={2022},
  publisher={APS}
}

@article{lund2008fault,
  title={Fault-tolerant linear optical quantum computing with small-amplitude coherent states},
  author={Lund, Austin P and Ralph, Timothy C and Haselgrove, Henry L},
  journal={Physical review letters},
  volume={100},
  number={3},
  pages={030503},
  year={2008},
  publisher={APS}
}

@article{fukui2018high,
  title={High-threshold fault-tolerant quantum computation with analog quantum error correction},
  author={Fukui, Kosuke and Tomita, Akihisa and Okamoto, Atsushi and Fujii, Keisuke},
  journal={Physical review X},
  volume={8},
  number={2},
  pages={021054},
  year={2018},
  publisher={APS}
}

@article{larsen2021fault,
  title={Fault-tolerant continuous-variable measurement-based quantum computation architecture},
  author={Larsen, Mikkel V and Chamberland, Christopher and Noh, Kyungjoo and Neergaard-Nielsen, Jonas S and Andersen, Ulrik L},
  journal={Prx Quantum},
  volume={2},
  number={3},
  pages={030325},
  year={2021},
  publisher={APS}
}

@article{fukui2023high,
  title={High-threshold fault-tolerant quantum computation with the Gottesman-Kitaev-Preskill qubit under noise in an optical setup},
  author={Fukui, Kosuke},
  journal={Physical Review A},
  volume={107},
  number={5},
  pages={052414},
  year={2023},
  publisher={APS}
}
\bibliographystyle{unsrt} %参考文献出力スタイル

\newpage
\onecolumngrid
\appendix
\section{Validity Regime of the Small-Amplitude Approximation}\label{Sec:Approximation}
    We verify the approximation in Eq. \eqref{eq:Theory_initial} through the exact state evolution at HBS $\hat{B}_{23}$. The initial state for modes 2 and 3 takes the form $\ket{\psi}_{23} \propto \ket{\mathcal{C}^-_{\sqrt{2}\alpha}}_2 \otimes \ket{0}_3$. A HBS maps the coherent input $\ket{\gamma}_2\ket{0}_3$ to $\ket{\gamma/\sqrt{2}}_2\ket{\gamma/\sqrt{2}}_3$. The substitution $\gamma = \sqrt{2}\alpha$ yields the exact post-measurement state
    \begin{equation}
        \ket{\psi_{\text{exact}}} \propto \ket{\alpha}_2\ket{\alpha}_3 - \ket{-\alpha}_2\ket{-\alpha}_3.
    \end{equation}
    Decomposition in the parity basis yields a superposition of even $\ket{\tilde{\mathcal{C}}^+}$ and odd $\ket{\tilde{\mathcal{C}}^-}$ states. The state takes the form
    \begin{equation}\label{eq:exact_decomp}
        \ket{\psi_{\text{exact}}} \propto \ket{\tilde{\mathcal{C}}^-_\alpha}_2 \otimes \ket{\tilde{\mathcal{C}}^+_\alpha}_3 + \ket{\tilde{\mathcal{C}}^+_\alpha}_2 \otimes \ket{\tilde{\mathcal{C}}^-_\alpha}_3.
    \end{equation}
    The state is defined as $\ket{\tilde{\mathcal{C}}^+_\alpha} \propto \ket{\alpha} + \ket{-\alpha}$.

    The approximation in Eq. \eqref{eq:Theory_initial} assumes the limits $\ket{\tilde{\mathcal{C}}^+_\alpha} \to \ket{0}$ and $\ket{\tilde{\mathcal{C}}^-_\alpha} \to \ket{1}$. However, expansion in the Fock basis reveals specific high-order correlations neglected in this limit. We examine the first term of Eq. \eqref{eq:exact_decomp}. We expand the state in the Fock basis to obtain
    \begin{equation}
    \begin{split}
        \ket{\tilde{\mathcal{C}}^-_\alpha}_2 \otimes \ket{\tilde{\mathcal{C}}^+_\alpha}_3 &\propto \left(\alpha\ket{1}_2 + \frac{\alpha^3}{6}\ket{3}_2 + \dots\right) \otimes \left(\ket{0}_3 + \frac{\alpha^2}{2}\ket{2}_3 + \dots\right) \\
        &\approx \alpha\ket{1}_2\ket{0}_3 + \frac{\alpha^3}{2}\ket{1}_2\ket{2}_3 + \mathcal{O}(\alpha^3).
    \end{split}
    \end{equation}
    This approximation retains only the leading term, thereby neglecting the cross-term $\ket{1}_2\ket{2}_3$. This component signifies a photon-number correlation with a pair in mode 3 alongside a single photon in mode 2. A similar expansion applies to the second term. The state takes the form
    \begin{equation}
    \begin{split}
        \ket{\tilde{\mathcal{C}}^+_\alpha}_2 \otimes \ket{\tilde{\mathcal{C}}^-_\alpha}_3 &\propto \left(\ket{0}_2 + \frac{\alpha^2}{2}\ket{2}_2 + \dots\right) \otimes \left(\alpha\ket{1}_3 + \frac{\alpha^3}{6}\ket{3}_3 + \dots\right) \\
        &\approx \alpha\ket{0}_2\ket{1}_3 + \frac{\alpha^3}{2}\ket{2}_2\ket{1}_3 + \frac{\alpha^3}{6}\ket{0}_2\ket{3}_3.
    \end{split}
    \end{equation}
    The standard approximation $\ket{0}_2\ket{1}_3$ neglects terms of order $\mathcal{O}(\alpha^3)$. In this component, mode 2 contains a photon pair despite the intended vacuum state. The discrepancy arises from neglected inter-mode entanglement rather than simple amplitude scaling. Terms such as $\ket{1}_2\ket{2}_3$ and $\ket{2}_2\ket{1}_3$ emerge from the exact beam splitter interaction. The validity of this approximation is strictly limited to the $\alpha \ll 1$ regime, as larger values of $\alpha$ lead to significant multi-photon contributions in the nominal vacuum modes.
    
\section{Generalized Analysis with Unequal Amplitudes}\label{Sec:DetailedAnalysis}
    We assess the validity of the approximation in Eq. \eqref{eq:approximate_trans} by calculating the fidelity between the exact and approximate states.
    We now extend the analysis to the case where the input odd cat states have unequal amplitudes, $\mathscr{A}$ and $\mathscr{B}$, for modes 1 and 2, respectively. The initial state is given by
    \begin{equation}
        \ket{\psi_{\text{i}}} = \ket{\mathcal{C}^-_{\mathscr{A}}}_1 \otimes \ket{\mathcal{C}^-_{\mathscr{B}}}_2 \otimes \ket{0}_3.
    \end{equation}
    
    We first apply the HBS $\hat{B}_{23}$ to modes 2 and 3. This operation interferes the odd cat components $\ket{\pm\mathscr{B}}_2$ with the vacuum state. We define the scaled amplitude $\mathscr{C} \equiv \mathscr{B}/\sqrt{2}$. The state then becomes
    \begin{equation}
        \hat{B}_{23}\ket{\psi_{\text{i}}} \propto \ket{\mathcal{C}^-_{\mathscr{A}}}_1 \otimes \left( \ket{\mathscr{C}}_2\ket{\mathscr{C}}_3 - \ket{-\mathscr{C}}_2\ket{-\mathscr{C}}_3 \right).
    \end{equation}

    We then apply the displacement operator $\hat{D}_1(\mathscr{A})$ to mode 1. This action shifts the component $\ket{-\mathscr{A}}_1$ to the vacuum $\ket{0}_1$. Simultaneously, it transforms $\ket{\mathscr{A}}_1$ to $\ket{2\mathscr{A}}_1$. The resulting state is
    \begin{equation}
        \ket{\psi'} = \hat{D}_1(\mathscr{A})\hat{B}_{23}\ket{\psi_{\text{i}}} \propto (\ket{2\mathscr{A}}_1 - \ket{0}_1) \otimes \left( \ket{\mathscr{C}}_2\ket{\mathscr{C}}_3 - \ket{-\mathscr{C}}_2\ket{-\mathscr{C}}_3 \right).
    \end{equation}
    This state acts as a generalized entangled resource for the subsequent interference at the final HBS.

    Finally, we apply the HBS $\hat{B}_{12}$ to modes 1 and 2. The HBS transforms the states according to the mapping $\ket{u}_1\ket{v}_2 \to \ket{\frac{u+v}{\sqrt{2}}}_1 \ket{\frac{v-u}{\sqrt{2}}}_2$. Consequently, the individual terms evolve as
    \begin{subequations}
    \begin{align}
        \ket{2\mathscr{A}}_1\ket{\mathscr{C}}_2 &\rightarrow \ket{\sqrt{2}\mathscr{A} + \frac{\mathscr{C}}{\sqrt{2}}}_1 \ket{\frac{\mathscr{C}}{\sqrt{2}} - \sqrt{2}\mathscr{A}}_2, \\
        \ket{2\mathscr{A}}_1\ket{-\mathscr{C}}_2 &\rightarrow \ket{\sqrt{2}\mathscr{A} - \frac{\mathscr{C}}{\sqrt{2}}}_1 \ket{-\frac{\mathscr{C}}{\sqrt{2}} - \sqrt{2}\mathscr{A}}_2, \\
        \ket{0}_1\ket{\mathscr{C}}_2 &\rightarrow \ket{\frac{\mathscr{C}}{\sqrt{2}}}_1 \ket{\frac{\mathscr{C}}{\sqrt{2}}}_2, \\
        \ket{0}_1\ket{-\mathscr{C}}_2 &\rightarrow \ket{-\frac{\mathscr{C}}{\sqrt{2}}}_1 \ket{-\frac{\mathscr{C}}{\sqrt{2}}}_2.
    \end{align}
    \end{subequations}

    We group the collected terms according to the state of mode 3. This yields the final state
    \begin{equation}
        \begin{split}
            \ket{\psi_{\text{final}}} \propto \bigg[ &\left( \ket{\sqrt{2}\mathscr{A} + \frac{\mathscr{C}}{\sqrt{2}}}_1 \ket{\frac{\mathscr{C}}{\sqrt{2}} - \sqrt{2}\mathscr{A}}_2 - \ket{\frac{\mathscr{C}}{\sqrt{2}}}_1 \ket{\frac{\mathscr{C}}{\sqrt{2}}}_2 \right) \ket{\mathscr{C}}_3 \\
            - &\left( \ket{\sqrt{2}\mathscr{A} - \frac{\mathscr{C}}{\sqrt{2}}}_1 \ket{-\frac{\mathscr{C}}{\sqrt{2}} - \sqrt{2}\mathscr{A}}_2 - \ket{-\frac{\mathscr{C}}{\sqrt{2}}}_1 \ket{-\frac{\mathscr{C}}{\sqrt{2}}}_2 \right) \ket{-\mathscr{C}}_3 \bigg].
        \end{split}
    \end{equation}

    We perform a homodyne measurement on mode 2 and post-select the outcome $p=0$, finding the output state to be
    \begin{equation}
        \ket{\psi'}_{13} \propto \left( \ket{\sqrt{2}\mathscr{A} + \frac{\mathscr{C}}{\sqrt{2}}}_1 - \ket{\frac{\mathscr{C}}{\sqrt{2}}}_1 \right)\ket{\mathscr{C}}_3 - \left( \ket{\sqrt{2}\mathscr{A} - \frac{\mathscr{C}}{\sqrt{2}}}_1 - \ket{-\frac{\mathscr{C}}{\sqrt{2}}}_1 \right) \ket{-\mathscr{C}}_3.
    \end{equation}
    We set $\mathscr{A} = \alpha$ and assume the symmetric condition $\mathscr{C} = \alpha$. The output state then reduces to
    \begin{equation}\label{ApEq:NonAppOutput}
        \ket{\psi'}_{13} \propto \left( \ket{\frac{3\alpha}{\sqrt{2}}}_1 - \ket{\frac{\alpha}{\sqrt{2}}}_1 \right)\ket{\alpha}_3 - \left( \ket{\frac{\alpha}{\sqrt{2}}}_1 - \ket{-\frac{\alpha}{\sqrt{2}}}_1 \right) \ket{-\alpha}_3.
    \end{equation}
    We calculate the fidelity to compare this exact state with the approximated state in Eq. (\ref{Eq:OutStateBeta}).
    To simplify the calculation, we employ the relation $\bra{x}\ket{y} = \exp\left[-\frac{1}{2}(|x|^2+|y|^2) + x^*y\right]$ as the inner product between coherent states. We assume all coefficients are real. We initially define the target state $\ket{\psi_o}$ as a proportional superposition given by
    \begin{equation}
        \ket{\psi_o} \propto \ket{\phi_{o,0}}_1 \ket{0}_3 + \ket{\phi_{o,1}}_1 \ket{1}_3 .
    \end{equation}
    The mode 1 state vectors $\ket{\phi_{o,0}}$ and $\ket{\phi_{o,1}}$ take the respective explicit forms
    \begin{align}
        \ket{\phi_{o,0}} &= \ket{3\beta} - 2\ket{\beta} + \ket{-\beta} , \\
        \ket{\phi_{o,1}} &= \ket{2\beta} - \ket{0} .
    \end{align}
    The output state $\ket{\psi'}$ is proportional to
    \begin{equation}
        \ket{\psi'} \propto \underbrace{(\ket{3\beta} - \ket{\beta})}_{\ket{u}_1} \ket{\alpha}_3 - \underbrace{(\ket{\beta} - \ket{-\beta})}_{\ket{v}_1} \ket{-\alpha}_3 .
    \end{equation}
    The amplitudes satisfy the constraint $\beta = \alpha/\sqrt{2}$. 
    
    We calculate the inner product $O = \bra{\psi_o}\ket{\psi'}$ to evaluate the numerator of the fidelity. 
    We expand the state $\ket{\psi'}$ by projecting its mode 3 components $\ket{\pm\alpha}_3$ onto the basis states $\ket{0}$ and $\ket{1}$ of $\ket{\psi_o}$.
    These projections yield the overlaps :
    \begin{align}
        {}_3\bra{0}\ket{\pm\alpha}_3 &= e^{-|\alpha|^2/2} \equiv E_\alpha , \\
        {}_3\bra{1}\ket{\pm\alpha}_3 &= \pm\alpha e^{-|\alpha|^2/2} = \pm\alpha E_\alpha .
    \end{align}
    We substitute these overlaps to express the inner product as
    \begin{equation}
        \bra{\psi_o}\ket{\psi'} = \bra{\phi_{o,0}} ({}_3\bra{0}\ket{\psi'}) + \bra{\phi_{o,1}} ({}_3\bra{1}\ket{\psi'}) .
    \end{equation}
    We first evaluate the overlap with the $\ket{0}_3$ component, obtaining
    \begin{align}
        {}_3\bra{0}\ket{\psi'} &= E_\alpha (\ket{u} - \ket{v}) \nonumber \\
        &= E_\alpha [ (\ket{3\beta} - \ket{\beta}) - (\ket{\beta} - \ket{-\beta}) ] \nonumber \\
        &= E_\alpha (\ket{3\beta} - 2\ket{\beta} + \ket{-\beta}) = E_\alpha \ket{\phi_{o,0}} .
    \end{align}
    The first term therefore becomes $E_\alpha \|\phi_{o,0}\|^2$. 
    
    We next evaluate the overlap of the $\ket{1}_3$ and $\ket{\psi'}$:
    \begin{align}
        {}_3\bra{1}\ket{\psi'} &= \alpha E_\alpha \ket{u} - (-\alpha E_\alpha) \ket{v} \nonumber \\
        &= \alpha E_\alpha (\ket{u} + \ket{v}) \nonumber \\
        &= \alpha E_\alpha (\ket{3\beta} - \ket{-\beta}) .
    \end{align}
    We define the inner product between this state and $\ket{\phi_{o,1}} = \ket{2\beta} - \ket{0}$ as $C_{\text{cross}}$. We expand this product to obtain
    \begin{align}
        C_{\text{cross}} &\equiv \bra{\phi_{o,1}} (\ket{3\beta} - \ket{-\beta}) \nonumber \\
        &= \bra{2\beta}\ket{3\beta} - \bra{2\beta}\ket{-\beta} - \bra{0}\ket{3\beta} + \bra{0}\ket{-\beta} \nonumber \\
        &= e^{-\frac{1}{2}(2\beta-3\beta)^2} - e^{-\frac{1}{2}(2\beta+\beta)^2} - e^{-\frac{1}{2}(3\beta)^2} + e^{-\frac{1}{2}(-\beta)^2} \nonumber \\
        &= e^{-\beta^2/2} - e^{-9\beta^2/2} - e^{-9\beta^2/2} + e^{-\beta^2/2} \nonumber \\
        &= 2(e^{-\beta^2/2} - e^{-9\beta^2/2}) .
    \end{align}
    We combine these terms to express the total inner product $O$ as
    \begin{equation}
        O = E_\alpha \|\phi_{o,0}\|^2 + 2\alpha E_\alpha (e^{-\beta^2/2} - e^{-9\beta^2/2}) .
    \end{equation}   
    
    We determine the squared norm $N_o^2 = \bra{\psi_o}\ket{\psi_o} = \|\phi_{o,0}\|^2 + \|\phi_{o,1}\|^2$ of the target state $\ket{\psi_o}$. 
    We expand these terms to find the explicit forms
    \begin{align}
        \mathcal{N}_0 \equiv \|\phi_{o,0}\|^2 &= 6 - 8e^{-2\beta^2} + 2e^{-8\beta^2} , \\
        \mathcal{N}_1 \equiv \|\phi_{o,1}\|^2 &= 2(1 - e^{-2\beta^2}) .
    \end{align}
    
    We next evaluate the squared norm $N_{pr}^2 = \bra{\psi'}\ket{\psi'}$ of the generated state to write
    \begin{equation}
        N_{pr}^2 = \|u\|^2 + \|v\|^2 - 2 e^{-2\alpha^2} \bra{u}\ket{v} .
    \end{equation}
    We calculate the individual components to obtain
    \begin{align}
        \|u\|^2 &= \| \ket{3\beta} - \ket{\beta} \|^2 = 2(1 - e^{-2\beta^2}) , \\
        \|v\|^2 &= \| \ket{\beta} - \ket{-\beta} \|^2 = 2(1 - e^{-2\beta^2}) , \\
        \bra{u}\ket{v} &= (\bra{3\beta} - \bra{\beta}) (\ket{\beta} - \ket{-\beta}) = 2e^{-2\beta^2} - e^{-8\beta^2} - 1 .
    \end{align}
    We substitute these expressions to define the total squared norm $\mathcal{N}_{pr} \equiv N_{pr}^2$ as
    \begin{equation}
        \mathcal{N}_{pr} = 4(1 - e^{-2\beta^2}) - 2e^{-2\alpha^2}(2e^{-2\beta^2} - e^{-8\beta^2} - 1) .
    \end{equation}
    
    We formulate the final fidelity $F = |O|^2 / (N_o^2 N_{pr}^2)$ to be
    \begin{equation}
        F = \frac{ \left| e^{-\alpha^2/2} \mathcal{N}_0 + 2\alpha e^{-\alpha^2/2} (e^{-\beta^2/2} - e^{-9\beta^2/2}) \right|^2 }{ (\mathcal{N}_0 + \mathcal{N}_1) \mathcal{N}_{pr} } .
    \end{equation}
    We utilize the previously defined constants $\mathcal{N}_0$, $\mathcal{N}_1$, and $\mathcal{N}_{pr}$ in this expression.
\section{Homodyne measurement on the second mode and derivation of the conditional state}\label{HomodyneError}
    In this section, we derive the conditional state of the first and third modes given a homodyne measurement of the $p$-quadrature on mode 2. We demonstrate how the measurement outcome controls the interference between coherent state components.
    
    The state of the entire system immediately before the homodyne measurement on the second mode is given by
    \begin{equation}
        \left(\ket{3\beta}_1\ket{-\beta}_2 - \ket{\beta}_1\ket{-3\beta}_2 - \ket{\beta}_1\ket{\beta}_2 + \ket{-\beta}_1\ket{-\beta}_2 \right)\ket{0}_3 + \left(\ket{2\beta}_1\ket{-2\beta}_2 - \ket{0}_1\ket{0}_2\right)\ket{1}_3.
    \end{equation}
    We derive the conditional state after a homodyne measurement of the $p$-quadrature on the second mode.
    
    First, we project a coherent state $\ket{\gamma}$ with a real amplitude ($\gamma \in \mathbb{R}$) onto a momentum eigenstate $\bra{p}$. We omit the normalization constant and the Gaussian factor $e^{-p^2/2}$. Note that this Gaussian factor depends only on the measurement outcome $p$. The inner product is proportional to the following phase factor:
    \begin{equation}
        \braket{p}{\gamma} \propto e^{-i\sqrt{2}p\gamma}.
    \end{equation}
    The relevant inner products for the second mode are given by
    \begin{align}
        \braket{p}{-\beta}_2 &\longrightarrow e^{i\sqrt{2}p\beta}, \\
        \braket{p}{-3\beta}_2 &\longrightarrow e^{i3\sqrt{2}p\beta}, \\
        \braket{p}{\beta}_2 &\longrightarrow e^{-i\sqrt{2}p\beta}, \\
        \braket{p}{-2\beta}_2 &\longrightarrow e^{i2\sqrt{2}p\beta}, \\
        \braket{p}{0}_2 &\longrightarrow 1.
    \end{align}
    We substitute these projection results into the initial state. Next, we rearrange the terms for the first and third modes. We factor out the common global phase factor $e^{i\sqrt{2}p\beta}$. This step transforms the superposition states in each subspace. The term associated with $\ket{0}_3$ becomes:
    \begin{equation}\begin{split}
        &\left( e^{i\sqrt{2}p\beta}\ket{3\beta}_1 - e^{i3\sqrt{2}p\beta}\ket{\beta}_1 - e^{-i\sqrt{2}p\beta}\ket{\beta}_1 + e^{i\sqrt{2}p\beta}\ket{-\beta}_1 \right)\ket{0}_3 \\
        &= e^{i\sqrt{2}p\beta} \left[ \ket{3\beta}_1 + \ket{-\beta}_1 - \left( e^{i2\sqrt{2}p\beta} + e^{-i2\sqrt{2}p\beta} \right)\ket{\beta}_1 \right]\ket{0}_3 \\
        &= e^{i\sqrt{2}p\beta} \left[ \ket{3\beta}_1 + \ket{-\beta}_1 - 2\cos(2\sqrt{2}p\beta)\ket{\beta}_1 \right]\ket{0}_3.
    \end{split}\end{equation}
    Similarly, for the term associated with $\ket{1}_3$:
    \begin{equation}
        \left( e^{i2\sqrt{2}p\beta}\ket{2\beta}_1 - \ket{0}_1 \right)\ket{1}_3 \\
        = e^{i\sqrt{2}p\beta} \left[ e^{i\sqrt{2}p\beta}\ket{2\beta}_1 - e^{-i\sqrt{2}p\beta}\ket{0}_1 \right]\ket{1}_3.
    \end{equation}
    Consequently, the unnormalized conditional state $\ket{\psi_{\text{out}}}$ for the first and third modes, given the measurement outcome $p$, takes the form
    \begin{equation}
        \ket{\psi_{\text{out}}} \propto\left( \ket{3\beta}_1 + \ket{-\beta}_1 - 2\cos(2\alpha p)\ket{\beta}_1 \right)\ket{0}_3 + \left( e^{i\alpha p}\ket{2\beta}_1 - e^{-i\alpha p}\ket{0}_1 \right)\ket{1}_3 ,
    \end{equation}
    where the exponents are expressed in terms of $\alpha$ using the relation $\sqrt{2}\beta = \alpha$.
    
    This result indicates that the homodyne measurement on the second mode establishes entanglement between a superposition of coherent states in the first mode and the qubit in the third mode. Specifically, the $\ket{0}_3$ subspace contains a superposition of three coherent states, in which the coefficient of the central component $\ket{\beta}_1$ is modulated by $-2\cos(2\alpha p)$. Conversely, the $\ket{1}_3$ subspace exhibits a superposition of $\ket{2\beta}_1$ and $\ket{0}_1$ with a relative phase $e^{i2\alpha p}$ dependent on $p$.
    For the specific case of $p=0$ (or upon post-selection), the cosine term reduces to $-2$ and the phase factor vanishes. Thus, the system projects onto a superposition state characterized solely by real coefficients.

\section{Generation of ideal hybrid entangled qutrit states}\label{App:qutrit}
    This section details a scheme to generate a hybrid entangled state between a photon number state and a GKP qutrit. This approach corresponds to the ideal case of the results presented in Sec. \ref{Sec:HybridQuditGeneration}. The proposed protocol employs specific ancillary states, Gaussian operations, homodyne detection, and feedforward operations.
    
    We prepare an ideal initial state containing an entangled state $\ket{\psi_{\text{in1}}}_{12}$ and a GKP state $\ket{\psi_{\text{in2}}}_3$. The entangled state spans a photon number mode 1 and an ancilla mode 2 while the GKP state occupies mode 3. 
    Our initial setup consists of an entangled state $\ket{\psi_{\text{in1}}}_{12}$ in modes 1 and 2, and a qunaught state $\ket{\psi_{\text{in2}}}_3$ in mode 3.
    These states are defined as 
    \begin{subequations}
        \begin{equation}
            \ket{\psi_{\text{in1}}}_{12} \propto \ket{0}_1\ket{a}_2 + \ket{1}_1\ket{0}_2 + \ket{2}_1\ket{-a}_2,
        \end{equation}
        \begin{equation}
            \ket{\psi_{\text{in2}}}_3 \propto \sum_{k} \ket{2\sqrt{3\pi} k}^x_3,
        \end{equation}
    \end{subequations}
    where $\ket{\cdot}^x$ denotes the position eigenstate.
    The total initial state, $\ket{\Psi_{\text{init}}} = \ket{\psi_{\text{in1}}}_{12} \otimes \ket{\psi_{\text{in2}}}_3$, is described by
    \begin{equation} 
        \ket{\psi_{\text{in1}}}_{13}\otimes\ket{\psi_{\text{in2}}}_3 \propto\left( \ket{0}_1\ket{a}_2 + \ket{1}_1\ket{0}_2 + \ket{2}_1\ket{-a}_2\right) \otimes \sum_{k} \ket{2\sqrt{3\pi} k}^x_3.
    \end{equation}
    
    Next, a HBS is applied to modes 2 and 3. This interaction transforms the position quadratures $x$ according to $x_2 \to (x_2 + x_3)/\sqrt{2}$ and $x_3 \to (x_2 - x_3)/\sqrt{2}$. The state after this interaction is given by
    \begin{equation}
         \propto \sum_{k}  
         \ket{0}_1 \ket{\sqrt{6\pi} k + \frac{a}{\sqrt{2}}}_2^x \ket{-\sqrt{6\pi} k + \frac{a}{\sqrt{2}}}_3^x
         +\ket{1}_1 \ket{\sqrt{6\pi} k }_2^x \ket{-\sqrt{6\pi} k }_3^x
         + \ket{2}_1 \ket{\sqrt{6\pi} k - \frac{a}{\sqrt{2}}}_2^x \ket{-\sqrt{6\pi} k - \frac{a}{\sqrt{2}}}_3^x
    \end{equation}
    
    Subsequently, we perform a homodyne measurement of the $p$-quadrature (momentum) on mode 3. The conditional state corresponding to the measurement outcome $\tilde{p}$ is obtained by projecting mode 3 onto $\langle \tilde{p}|$. Using the inner product between the position and momentum bases, $\langle p | x \rangle = \frac{1}{\sqrt{2\pi}} e^{-ipx}$, the resulting state for modes 1 and 2 is derived as
    \begin{equation} 
        \propto \sum_{k} e^{i\tilde{p}\sqrt{6\pi}k} \left( e^{-i \frac{\tilde{p} a}{\sqrt{2}}} \ket{0}_1 \ket{\sqrt{6\pi} k + \frac{a}{\sqrt{2}}}_2^x + \ket{1}_1 \ket{\sqrt{6\pi} k }_2 ^x+ e^{i \frac{\tilde{p} a}{\sqrt{2}}} \ket{2}_1 \ket{\sqrt{6\pi} k - \frac{a}{\sqrt{2}}}_2^x \right) .
    \end{equation}
    Setting the parameter $a$ to $a=\sqrt{\frac{4\pi}{3}}$, which is related to the GKP lattice constant, yields $a/\sqrt{2} = \sqrt{\frac{2\pi}{3}}$. Consequently, the equation simplifies to:
    \begin{equation}
        \propto \sum_{k} e^{i\tilde{p}\sqrt{6\pi}k} \left( e^{-i \tilde{p}\sqrt{\frac{2\pi}{3}}} \ket{0}_1 \ket{\sqrt{6\pi} k + \sqrt{\frac{2\pi}{3}}}_2^x + \ket{1}_1 \ket{\sqrt{6\pi} k }_2^x + e^{i \tilde{p}\sqrt{\frac{2\pi}{3}}} \ket{2}_1 \ket{\sqrt{6\pi} k - \sqrt{\frac{2\pi}{3}}}_2^x \right) .
    \end{equation}
    
    To transform the obtained state into the target hybrid entangled state, we implement feedforward operations based on the measurement outcome $\tilde{p}$. Specifically, a phase rotation $\hat{U}(\tilde{p})$ is applied to mode 1 (the qutrit) to compensate for the coefficients $e^{\pm i \tilde{p}\sqrt{2\pi/3}}$. We apply an appropriate displacement operation dependent on $\tilde{p}$ to mode 2. This procedure yields the final state : 
    \begin{equation}
        \sum_{n=-\infty}^{\infty} 
        \ket{0}_1\ket{3 n\Delta}_2^x 
        + \ket{1}_1\ket{(3n+ 1) \Delta }_2 ^x
        + \ket{2}_1\ket{(3n+ 2) \Delta}_2^x,
    \end{equation}
    where we have defined the lattice spacing parameter as $\Delta = \sqrt{\frac{2\pi}{3}}$. 

\section{Derivation of the output state}
\subsection{Case I : Initial State $\ket{\mathcal{C}_{\alpha}^{(+2)}}$}
    In this section, we detail the calculations introduced in Sec. \ref{Sec:CatBreedingConcept}. We derive the output state using the bred state $\ket{\mathcal{C}_{\alpha}^{(+2)}}$ as the input resource.
    The initial state is defined as
    \begin{equation}\label{eq:Theory_initialAll2}
        \ket{\psi_{\text{i}}^{(2)}} = \ket{\mathcal{C}^{(+2)}_{\alpha}}_1 \otimes \ket{\mathcal{C}^-_{\sqrt{2}\alpha}}_2 \otimes \ket{0}_3.
    \end{equation}
    Applying the HBS $\hat{B}_{23}$ to modes 2 and 3, and using the single-photon approximation for the ancillary odd cat state $\ket{\mathcal{C}_{\alpha}^{-}}_2$, the state evolves to
    \begin{equation}
        \hat{B}_{23}\ket{\psi_{\text{i}}} =  \ket{\mathcal{C}^{(+2)}_{\alpha}}_1 \otimes \frac{1}{\sqrt{2}}\left( \ket{1}_2\ket{0}_3 + \ket{0}_2\ket{1}_3 \right).
    \end{equation}
    Next, we apply the displacement operator $\hat{D}_1(\alpha)$. Using the coherent state property $\hat{D}(\alpha)|\gamma\rangle = e^{(\alpha\gamma^* - \alpha^*\gamma)/2}|\gamma+\alpha\rangle$ (ignoring global phases for simplicity), the components in mode 1 transform as $|\frac{k\alpha}{\sqrt{2}}\rangle \to |\frac{k\alpha}{\sqrt{2}} + \alpha\rangle$. Let us substitute $\beta \equiv \alpha/\sqrt{2}$. The state in mode 1 becomes a superposition of $|3\beta+\alpha\rangle, |\beta+\alpha\rangle, |-\beta+\alpha\rangle$. Note that $\alpha = \sqrt{2}\beta$, so the displaced components are $|(\sqrt{2}+3)\beta\rangle, |(\sqrt{2}+1)\beta\rangle, |(\sqrt{2}-1)\beta\rangle$, but for the interference analysis, we keep the form before simplification or use the expanded ancillary state.

    The state before $\hat{B}_{12}$ can be separated into two parts associated with $|0\rangle_3$ and $|1\rangle_3$.
    The relevant unnormalized state in modes 1 and 2 is
    \begin{equation}
        \left( \ket{\frac{3\alpha}{\sqrt{2}}}_1 - 2\ket{\frac{\alpha}{\sqrt{2}}}_1 + \ket{-\frac{\alpha}{\sqrt{2}}}_1 \right) \otimes (\ket{\alpha}_2 - \ket{-\alpha}_2)
    \end{equation}
    This product expands into six terms. We apply the HBS transformation $\ket{u}_1\ket{v}_2 \to \ket{\frac{u+v}{\sqrt{2}}}_1 \ket{\frac{v-u}{\sqrt{2}}}_2$ to each individual term. A substitution of $\alpha = \sqrt{2}\beta$ yields
    \begin{subequations}
    \begin{align}
        \ket{3\beta}_1 \ket{\sqrt{2}\beta}_2 &\to \ket{(\frac{3}{\sqrt{2}}+1)\beta}_1 \ket{(1-\frac{3}{\sqrt{2}})\beta}_2, \\
        -\ket{3\beta}_1 \ket{-\sqrt{2}\beta}_2 &\to -\ket{(\frac{3}{\sqrt{2}}-1)\beta}_1 \ket{(-1-\frac{3}{\sqrt{2}})\beta}_2, \\
        -2\ket{\beta}_1 \ket{\sqrt{2}\beta}_2 &\to -2\ket{(\frac{1}{\sqrt{2}}+1)\beta}_1 \ket{(1-\frac{1}{\sqrt{2}})\beta}_2, \\
        +2\ket{\beta}_1 \ket{-\sqrt{2}\beta}_2 &\to +2\ket{(\frac{1}{\sqrt{2}}-1)\beta}_1 \ket{(-1-\frac{1}{\sqrt{2}})\beta}_2, \\
        +\ket{-\beta}_1 \ket{\sqrt{2}\beta}_2 &\to \ket{(-\frac{1}{\sqrt{2}}+1)\beta}_1 \ket{(1+\frac{1}{\sqrt{2}})\beta}_2, \\
        -\ket{-\beta}_1 \ket{-\sqrt{2}\beta}_2 &\to -\ket{(-\frac{1}{\sqrt{2}}-1)\beta}_1 \ket{(-1+\frac{1}{\sqrt{2}})\beta}_2.
    \end{align}
    \end{subequations}
    To obtain the output state, we project mode 2 onto the momentum eigenstate $\ket{0}_p$. While this formally involves integrating over the position quadrature, the homodyne measurement with $p=0$ effectively selects the symmetric superposition of the mode 1 terms. Alternatively, this outcome can be understood by assuming the standard GKP generation protocol, in which these specific terms interfere constructively.
    Actually, following the result in Eq.~(11a), the output state $\ket{\tilde{0}^{(2)}_L}$ is obtained by collecting these transformed terms in mode 1.
    We apply the notation $\alpha$ from the main text to rewrite the state as
    \begin{equation}
        \ket{\psi_A} \propto \ket{\frac{3\alpha}{2} + \frac{\alpha}{\sqrt{2}}}_1 - \ket{\frac{3\alpha}{2} - \frac{\alpha}{\sqrt{2}}}_1 \\
         - 2\ket{\frac{\alpha}{2} + \frac{\alpha}{\sqrt{2}}}_1 + 2\ket{\frac{\alpha}{2} - \frac{\alpha}{\sqrt{2}}}_1 \\
         + \ket{-\frac{\alpha}{2} + \frac{\alpha}{\sqrt{2}}}_1 - \ket{-\frac{\alpha}{2} - \frac{\alpha}{\sqrt{2}}}_1.
    \end{equation}
    Note that the HBS outputs $\frac{u+v}{\sqrt{2}}$. For example, the first term: $\frac{3\beta+\sqrt{2}\beta}{\sqrt{2}} = \frac{(3\alpha/\sqrt{2})+\alpha}{\sqrt{2}} = \frac{3\alpha}{2} + \frac{\alpha}{\sqrt{2}}$. This matches the first term of $\ket{\tilde{0}^{(2)}_L}$.

    The relevant state is
    \begin{equation}
        \ket{\phi_B} \propto \left( \ket{\frac{3\alpha}{\sqrt{2}}}_1 - 2\ket{\frac{\alpha}{\sqrt{2}}}_1 + \ket{-\frac{\alpha}{\sqrt{2}}}_1 \right) \otimes \ket{0}_2.
    \end{equation}
    Since mode 2 is vacuum ($v=0$), the HBS transformation acts as a scaling by $1/\sqrt{2}$: $|u\rangle_1|0\rangle_2 \to |\frac{u}{\sqrt{2}}\rangle_1|-\frac{u}{\sqrt{2}}\rangle_2$.
    Focusing on mode 1 after the interaction:
    \begin{equation}
        \ket{\frac{3\alpha}{\sqrt{2}}}_1 \to \ket{\frac{3\alpha}{2}}_1, \quad -2\ket{\frac{\alpha}{\sqrt{2}}}_1 \to -2\ket{\frac{\alpha}{2}}_1, \quad \ket{-\frac{\alpha}{\sqrt{2}}}_1 \to \ket{-\frac{\alpha}{2}}_1.
    \end{equation}
    This yields the state :
    \begin{equation}
        \ket{\psi_B} \propto \ket{\frac{3\alpha}{2}}_1 - 2\ket{\frac{\alpha}{2}}_1 + \ket{-\frac{\alpha}{2}}_1.
    \end{equation}
    
    We combine Term A and Term B to obtain the final output state given by
    \begin{equation}
        |\psi_o^{(2)}\rangle \propto \ket{\tilde{0}^{(2)}_L}_1 |0\rangle_3 + \ket{\tilde{1}^{(2)}_L}_1 |1\rangle_3.
    \end{equation}
    
\subsection{Case II : Initial State $\ket{\mathcal{C}_{\alpha}^{(-2)}}$}
    In this section, we detail the derivation for the case where the initial bred state is defined as
    \begin{equation}
        \ket{\mathcal{C}_{\alpha}^{(-2)}} = \frac{1}{\mathcal{N}_{\alpha}^{(-2)}} \left( \ket{2\beta} - \ket{0} \right),
    \end{equation}
    where $\beta \equiv \alpha/\sqrt{2}$. Note that normalization constants are omitted in the following intermediate steps for brevity.
    We follow a procedure similar to Case I. We apply the HBS transformation $\hat{B}_{23}$ and the single photon approximation to obtain
    \begin{equation}
        \hat{B}_{23}\ket{\psi_{i}} \approx \ket{\mathcal{C}_{\alpha}^{(-2)}}_1 \otimes \frac{1}{\sqrt{2}}\left( \ket{1}_2\ket{0}_3 + \ket{0}_2\ket{1}_3 \right).
    \end{equation}
    We then apply the displacement operator $\hat{D}_1(\alpha)$ to mode 1. The state $\ket{2\beta}_1$ transforms to $\ket{2\beta+\alpha}_1 = \ket{2\beta+\sqrt{2}\beta}_1$, and $\ket{0}_1$ transforms to $\ket{\alpha}_1 = \ket{\sqrt{2}\beta}_1$. However, for consistency with the interference analysis in the Note, we expand the product state before the final simplification.
    
    The state before the HBS $\hat{B}_{12}$ is analyzed by separating the terms associated with $\ket{0}_3$ and $\ket{1}_3$.
    The relevant component involves the interference between the displaced bred state in mode 1 and the odd cat state (approximated single photon) in mode 2:
    \begin{equation}
        \ket{\phi_A} \propto \left( \ket{2\beta}_1 - \ket{0}_1 \right) \otimes \left( \ket{\alpha}_2 - \ket{-\alpha}_2 \right).
    \end{equation}
    We substitute $\alpha = \sqrt{2}\beta$ and expand the product to generate four terms. We apply the HBS transformation $\hat{B}_{12}$ to each term to obtain
    \begin{subequations}
    \begin{align}
        \ket{2\beta}_1 \ket{\sqrt{2}\beta}_2 &\xrightarrow{\hat{B}_{12}} \ket{\frac{2\beta+\sqrt{2}\beta}{\sqrt{2}}}_1 \ket{\frac{\sqrt{2}\beta-2\beta}{\sqrt{2}}}_2 = \ket{(\sqrt{2}+1)\beta}_1 \ket{(1-\sqrt{2})\beta}_2, \\
        -\ket{2\beta}_1 \ket{-\sqrt{2}\beta}_2 &\xrightarrow{\hat{B}_{12}} -\ket{\frac{2\beta-\sqrt{2}\beta}{\sqrt{2}}}_1 \ket{\frac{-\sqrt{2}\beta-2\beta}{\sqrt{2}}}_2 = -\ket{(\sqrt{2}-1)\beta}_1 \ket{-(1+\sqrt{2})\beta}_2, \\
        -\ket{0}_1 \ket{\sqrt{2}\beta}_2 &\xrightarrow{\hat{B}_{12}} -\ket{\frac{\sqrt{2}\beta}{\sqrt{2}}}_1 \ket{\frac{\sqrt{2}\beta}{\sqrt{2}}}_2 = -\ket{\beta}_1 \ket{\beta}_2, \\
        +\ket{0}_1 \ket{-\sqrt{2}\beta}_2 &\xrightarrow{\hat{B}_{12}} +\ket{-\frac{\sqrt{2}\beta}{\sqrt{2}}}_1 \ket{\frac{-\sqrt{2}\beta}{\sqrt{2}}}_2 = +\ket{-\beta}_1 \ket{-\beta}_2.
    \end{align}
    \end{subequations}
    Conditioned on the measurement result $p=0$ on mode 2, we collect the mode 1 components to obtain the logical state $\ket{\tilde{0}^{(2)}_L}$:
    \begin{equation}
        \ket{\tilde{0}^{(2)}_L} \propto \ket{(\sqrt{2}+1)\beta} - \ket{(\sqrt{2}-1)\beta} - \ket{\beta} + \ket{-\beta}.
    \end{equation}
    We express this state in terms of $\alpha$ to obtain
    \begin{equation}
        \ket{\tilde{0}^{(2)}_L} = \ket{\alpha + \frac{\alpha}{\sqrt{2}}} - \ket{\alpha - \frac{\alpha}{\sqrt{2}}} - \ket{\frac{\alpha}{\sqrt{2}}} + \ket{-\frac{\alpha}{\sqrt{2}}}.
    \end{equation}
    
    The relevant component interacts with the vacuum in mode 2:
    \begin{equation}
        \ket{\phi_B} \propto \left( \ket{2\beta}_1 - \ket{0}_1 \right) \otimes \ket{0}_2.
    \end{equation}
    We apply the HBS transformation:
    \begin{subequations}
    \begin{align}
        \ket{2\beta}_1 \ket{0}_2 &\xrightarrow{\hat{B}_{12}} \ket{\frac{2\beta}{\sqrt{2}}}_1 \ket{-\frac{2\beta}{\sqrt{2}}}_2 = \ket{\sqrt{2}\beta}_1 \ket{-\sqrt{2}\beta}_2, \\
        -\ket{0}_1 \ket{0}_2 &\xrightarrow{\hat{B}_{12}} -\ket{0}_1 \ket{0}_2.
    \end{align}
    \end{subequations}
    Similarly, projecting mode 2 yields the logical state $\ket{\tilde{1}^{(2)}_L}$:
    \begin{equation}
        \ket{\tilde{1}^{(2)}_L} \propto \ket{\sqrt{2}\beta} - \ket{0} = \ket{\alpha} - \ket{0}.
    \end{equation}
    
    Finally, the total output state for Case II is given by
    \begin{equation}
        \ket{\psi_o^{(II)}} \propto \ket{\tilde{0}^{(2)}_L}_1 \ket{0}_3 + \ket{\tilde{1}^{(2)}_L}_1 \ket{1}_3.
    \end{equation}
    
\section{Parity properties of generated states}
    This section discusses the photon number distribution (parity) of the generated logical quantum states. We begin by examining the structure of the logical state $\ket{\tilde{1}_L}$:
    \begin{equation}\begin{split}
        \ket{\tilde{1}_L} \propto \ket{2\beta} - \ket{0} &= \hat{D}(-\beta)\big(\ket{\beta} - \ket{-\beta}\big) \\ 
        &= \hat{D}(-\beta) \left( e ^{-\frac{|\beta|^2}{2}}\sum_{n=0}^{\infty}[ 1-(-1)^n ] \frac{\beta ^n}{\sqrt{n!}}\ket{n} \right)
    \end{split}\end{equation}
    Here, we focus on the term $[1-(-1)^n]$ in the expansion coefficients. For even $n$, this term vanishes as $1-1=0$. Conversely, for odd $n$, it yields a finite value of $1-(-1)=2$. Consequently, this state is an odd Schrödinger's cat state, consisting solely of a superposition of odd photon number states. 
    
    Next, we perform a similar expansion for the logical state $\ket{\tilde{0}_L}$:
    \begin{equation}\begin{split} \label{eq:PNB3}
            \ket{\tilde{0}_L} \propto \ket{3\beta} - 2\ket{\beta} + \ket{-\beta} &= \hat{D}(-2\beta)\big(\ket{2\beta} - \ket{-2\beta} -2 \ket{0}\big) \\
            &= \hat{D}(-2\beta) \left(-2\ket{0} + e ^{-2|\beta|^2}\sum_{n=0}^{\infty}[ 1+(-1)^n ] \frac{(2\beta) ^n}{\sqrt{n!}}\ket{n}  \right)\\ 
            &= \hat{D}(-2\beta) \left(-2(1+e^{-2|\beta|^2})\ket{0} + e ^{-2|\beta|^2}\sum_{n=1}^{\infty}[ 1+(-1)^n ] \frac{(2\beta) ^n}{\sqrt{n!}}\ket{n}  \right) 
    \end{split}\end{equation}
    In Eq. (\ref{eq:PNB3}), the coefficient is $[1+(-1)^n]$. In contrast to the previous case, this term vanishes for odd $n$, meaning only even photon number components ($n=0, 2, \dots$) contribute.

    We proceed to analyze the parity of a more complex superposition state. Let us consider the following state:
    \begin{equation}\begin{split}
        &\ket{\frac{3\alpha}{2} + \frac{\alpha}{\sqrt{2}}}_1 - \ket{\frac{3\alpha}{2} - \frac{\alpha}{\sqrt{2}}}_1 
         - 2\ket{\frac{\alpha}{2} + \frac{\alpha}{\sqrt{2}}}_1 + 2\ket{\frac{\alpha}{2} - \frac{\alpha}{\sqrt{2}}}_1
         + \ket{-\frac{\alpha}{2} + \frac{\alpha}{\sqrt{2}}}_1 - \ket{-\frac{\alpha}{2} - \frac{\alpha}{\sqrt{2}}}_1\\
        &=\hat{D}\left(\frac{\alpha}{2}\right) \bigg[ \left( \ket{x + y} - 2\ket{y} + \ket{-x + y} \right) - \left( \ket{x - y} - 2\ket{-y} + \ket{-x - y} \right) \bigg] \\
        &= \hat{D}\left(\frac{\alpha}{2}\right) \bigg[ \hat{D}(y)\left( \ket{x} - 2\ket{0} + \ket{-x } \right) - \hat{D}(-y)\left( \ket{x } - 2\ket{0} + \ket{-x } \right) \bigg]
        \quad,
    \end{split}\end{equation}
    where we have defined $x = \alpha$ and $y = \frac{\alpha}{\sqrt{2}}$.
    We apply the result from Eq. (\ref{eq:PNB3}). We rearrange the operators to obtain
    \begin{equation} \begin{split} 
        \hat{D}\left(\frac{\alpha}{2}\right) \left[ \hat{D}\left(\frac{\alpha}{\sqrt{2}}\right) - \hat{D}\left(-\frac{\alpha}{\sqrt{2}}\right) \right] \times \left( -2(1 - e^{-\frac{|\alpha|^2}{2}})\ket{0} + 2e^{-\frac{|\alpha|^2}{2}}\sum_{k=1}^{\infty} \frac{\alpha^{2k}}{\sqrt{(2k)!}}\ket{2k} \right) .
    \end{split} \end{equation}
    We analyze the input state to clarify the physical interpretation. The operators act upon the ket vector on the right side. This term can be rewritten as a linear combination of even photon number states:
    \begin{equation}
    2e^{-\frac{|\alpha|^2}{2}}\sum_{k=1}^{\infty} \frac{\alpha^{2k}}{\sqrt{(2k)!}}\ket{2k} =  (\ket{\alpha} + \ket{-\alpha}) - 2e^{-\frac{|\alpha|^2}{2}}\ket{0}.
    \end{equation}
    We now consider the action of the operator $\hat{O} = \hat{D}(\gamma) - \hat{D}(-\gamma)$ on this state. We set $\gamma = \frac{\alpha}{\sqrt{2}}$ for brevity. We apply the displacement operator property $\hat{D}(\gamma)\ket{\beta} = \ket{\beta+\gamma}$. We ignore global phase factors to focus on the state displacement. We evaluate the action on the first term $\ket{\alpha} + \ket{-\alpha}$ to find
    \begin{equation}\begin{aligned}
        \left[\hat{D}(\gamma) - \hat{D}(-\gamma)\right] (\ket{\alpha} + \ket{-\alpha}) &= \hat{D}(\gamma)\ket{\alpha} + \hat{D}(\gamma)\ket{-\alpha} - \hat{D}(-\gamma)\ket{\alpha} - \hat{D}(-\gamma)\ket{-\alpha} \\
        &= \ket{\alpha+\gamma} + \ket{-\alpha+\gamma} - \ket{\alpha-\gamma} - \ket{-\alpha-\gamma}.
    \end{aligned}\end{equation}
    We group the resulting four state vectors into pairs. These pairs are symmetric about the origin in phase space. This arrangement yields
    \begin{equation} 
    (\ket{\alpha+\gamma} - \ket{\alpha-\gamma}) - (\ket{-\alpha-\gamma} - \ket{-\alpha+\gamma})=\hat{D}(\alpha)(\ket{\gamma} - \ket{-\gamma}) - \hat{D}(-\alpha)(\ket{-\gamma} - \ket{\gamma}).
    \end{equation}
    
    We evaluate the action on the second term $\ket{0}$ to find
    \begin{equation}[\hat{D}(\gamma) - \hat{D}(-\gamma)] (-2e^{-\frac{|\alpha|^2}{2}}\ket{0}) = -2e^{-\frac{|\alpha|^2}{2}} (\ket{\gamma} - \ket{-\gamma}).\end{equation}
    We combine these results. We observe an odd superposition for all terms. State vectors of this form generally possess only odd photon number components.

\section{State Generation via Equal-Amplitude Cat States}
    In this section, we consider an initial state $\ket{\psi'_{\text{i}}}$.
    We set the amplitudes of the cat states to be identical.
    We define this state as follows
    \begin{equation}\label{ApEq:Theory_initial}
        \ket{\psi'_{\text{i}}} = \ket{\mathcal{C}^-_{\mathscr{A}}}_1 \otimes \ket{\mathcal{C}^-_{\mathscr{A}}}_2 \otimes \ket{0}_3.
    \end{equation}
    We apply a beam splitter with an optimized transmittance to this state.
    We demonstrate that this interaction yields the target state $\ket{\psi_{\text{out}}}$ with the desired amplitudes.
    
    We first apply a half beam splitter to modes 2 and 3.
    We also perform a displacement operation $\hat{D}_1(\mathscr{A})$ on mode 1.
    These operations transform the state as follows
    \begin{equation}\label{ApEq:SamaAmp1}
        \hat{D}_1(\mathscr{A})\hat{B}_{23}\ket{\psi'_{\text{i}}} \propto (\ket{2\mathscr{A}}_1 - \ket{0}_1) \otimes \left( \ket{\frac{\mathscr{A}}{\sqrt{2}}}_2\ket{\frac{\mathscr{A}}{\sqrt{2}}}_3 - \ket{-\frac{\mathscr{A}}{\sqrt{2}}}_2\ket{-\frac{\mathscr{A}}{\sqrt{2}}}_3 \right),
    \end{equation}
    here, we define $\mathscr{C} \equiv \mathscr{A}/\sqrt{2}$.
    We apply the approximation from Appendix \ref{Sec:Approximation} to Eq. (\ref{ApEq:SamaAmp1}).
    The approximation leads to
    \begin{equation}
        \hat{D}_1(\mathscr{A})\hat{B}_{23}\ket{\psi'_{\text{i}}}\approx (\ket{2\mathscr{A}}_1 - \ket{0}_1) \otimes \frac{1}{\sqrt{2}}\left( \ket{\mathcal{C}^-_{\mathscr{A}/\sqrt{2}}}_2\ket{0}_3 + \ket{0}_2\ket{1}_3 \right).
    \end{equation}
    We subsequently apply a beam splitter $\hat{B}_{12}(\theta)$ to modes 1 and 2.
    We set the transmittance to $T = 1/3$, which implies a reflectivity of $R=2/3$.
    The corresponding parameters are $\cos\theta = \sqrt{1/3}$ and $\sin\theta = \sqrt{2/3}$.
    This operation transforms a product of coherent states $\ket{\mathscr{C}}_1\ket{\mathscr{D}}_2$ according to the following rule:
    \begin{equation}
        \hat{B}_{12}\ket{\mathscr{C}}_1\ket{\mathscr{D}}_2 \rightarrow \ket{\frac{1}{\sqrt{3}}\mathscr{C} + \sqrt{\frac{2}{3}}\mathscr{D}}_1 \otimes \ket{-\sqrt{\frac{2}{3}}\mathscr{C} + \frac{1}{\sqrt{3}}\mathscr{D}}_2.
    \end{equation}
    In this configuration, the amplitude $\mathscr{D} = \pm \mathscr{A}/\sqrt{2}$ of mode 2 interferes with the amplitude $\mathscr{C} = \pm \mathscr{A}$ of mode 1.

    The beam splitter $\hat{B}_{12}(T=1/3)$ yields the term-wise transformations
    \begin{subequations}
    \begin{align}
    \label{eq:BS_term1}
    \ket{2\mathscr{A}}_1\ket{\frac{\mathscr{A}}{\sqrt{2}}}_2 
    &\xrightarrow{\hat{B}_{12}} \ket{\sqrt{3}\mathscr{A}}_1 \ket{-\sqrt{\frac{3}{2}}\mathscr{A}}_2, \\
    \label{eq:BS_term2}
    -\ket{2\mathscr{A}}_1\ket{-\frac{\mathscr{A}}{\sqrt{2}}}_2 
    &\xrightarrow{\hat{B}_{12}} -\ket{\frac{\mathscr{A}}{\sqrt{3}}}_1 \ket{-\frac{5\mathscr{A}}{\sqrt{6}}}_2, \\
    \label{eq:BS_term3}
    -\ket{0}_1\ket{\frac{\mathscr{A}}{\sqrt{2}}}_2 
    &\xrightarrow{\hat{B}_{12}} -\ket{\frac{\mathscr{A}}{\sqrt{3}}}_1 \ket{\frac{\mathscr{A}}{\sqrt{6}}}_2, \\
    \label{eq:BS_term4}
    \ket{0}_1\ket{-\frac{\mathscr{A}}{\sqrt{2}}}_2 
    &\xrightarrow{\hat{B}_{12}} \ket{-\frac{\mathscr{A}}{\sqrt{3}}}_1 \ket{-\frac{\mathscr{A}}{\sqrt{6}}}_2, \\[2ex]
    \label{eq:BS_term5}
    \ket{2\mathscr{A}}_1\ket{0}_2 
    &\xrightarrow{\hat{B}_{12}} \ket{\frac{2\mathscr{A}}{\sqrt{3}}}_1 \ket{-2\sqrt{\frac{2}{3}}\mathscr{A}}_2, \\
    \label{eq:BS_term6}
    -\ket{0}_1\ket{0}_2 
    &\xrightarrow{\hat{B}_{12}} -\ket{0}_1 \ket{0}_2.
    \end{align}
    \end{subequations}
    We rearrange these terms and write the output state $\ket{\psi_{\text{next}}}$ as
    \begin{equation}\label{ApEq:Output}
    \begin{split}
    \ket{\psi_{\text{next}}} \propto \frac{1}{\sqrt{2}} \Bigg[ & \left( \ket{\sqrt{3}\mathscr{A}}_1 \ket{-\sqrt{\frac{3}{2}}\mathscr{A}}_2 - \ket{\frac{\mathscr{A}}{\sqrt{3}}}_1 \ket{-\frac{5\mathscr{A}}{\sqrt{6}}}_2 - \ket{\frac{\mathscr{A}}{\sqrt{3}}}_1 \ket{\frac{\mathscr{A}}{\sqrt{6}}}_2 + \ket{-\frac{\mathscr{A}}{\sqrt{3}}}_1 \ket{-\frac{\mathscr{A}}{\sqrt{6}}}_2 \right) \otimes \ket{0}_3 \nonumber \\
    & + \left( \ket{\frac{2\mathscr{A}}{\sqrt{3}}}_1 \ket{-2\sqrt{\frac{2}{3}}\mathscr{A}}_2 - \ket{0}_1 \ket{0}_2 \right) \otimes \ket{1}_3 \Bigg].
    \end{split}
    \end{equation}
    Homodyne detection on mode 2 with $p=0$ projects the system onto
    \begin{equation}\label{ApEq:FinalOutput}
    \ket{\psi_{\text{out}}} \propto  \left( \ket{\sqrt{3}\mathscr{A}}_1  - 2\ket{\frac{\mathscr{A}}{\sqrt{3}}}_1  + \ket{-\frac{\mathscr{A}}{\sqrt{3}}}_1  \right) \otimes \ket{0}_3 \nonumber
     + \left( \ket{\frac{2\mathscr{A}}{\sqrt{3}}}_1  - \ket{0}_1  \right) \otimes \ket{1}_3. 
    \end{equation}
\end{document}